\documentclass[
 reprint,
 superscriptaddress,
amsmath,amssymb,
aps,
prb,
floatfix,
]{revtex4-2}
\usepackage{hyperref}
    \hypersetup{colorlinks, citecolor=black, linkcolor= black, urlcolor=blue}
\usepackage{nameref}
\usepackage{graphicx}
\usepackage{dcolumn}
\usepackage{bm}
\usepackage{pgf}
\usepackage{physics}
\usepackage{dsfont}
\usepackage{xcolor}
\newcommand{\half}[1]{\frac{\mathbf{#1}}{2}}
\newcommand{\mvcm}{\frac{\text{MV}}{\text{cm}}}
\begin{document}
\preprint{APS/123-QED}
\title{Theory of Non-Linear Electron Relaxation in Thin Gold Films and Their Signatures in Optical Observables}
\author{Jonas Grumm}
\email{j.grumm@tu-berlin.de}
\affiliation{Nichtlineare Optik und Quantenelektronik, Institut für Physik und Astronomie, Technische Universität Berlin, 10623 Berlin, Germany}
\author{Malte Selig}
\affiliation{Nichtlineare Optik und Quantenelektronik, Institut für Physik und Astronomie, Technische Universität Berlin, 10623 Berlin, Germany}
\author{Holger Lange}
\affiliation{Institut für Physik und Astronomie, Universität Potsdam, 14476 Potsdam, Germany}
\author{Andreas Knorr}
\email{andreas.knorr@tu-berlin.de}
\affiliation{Nichtlineare Optik und Quantenelektronik, Institut für Physik und Astronomie, Technische Universität Berlin, 10623 Berlin, Germany}
\date{\today}
\begin{abstract}
Based on the momentum-resolved Boltzmann equation, we provide self-consistent numerical calculations of the dynamics of conduction electrons in thin noble metal films after linear and non-linear optical excitations with infrared and terahertz frequencies. Focusing exclusively on electron-phonon interaction, orientational relaxation is introduced and acts as dephasing of the optical excitation on a scale of tens of fs.
In the linear regime, our numerical results agree with the experimental fits to a Drude model and predicts for non-linear excitations a field strength dependency of the orientational relaxation rate. In the THz regime, where the orientational relaxation proceeds faster than the oscillation cycle of the excitation THz field, a new high order dissipative Kerr-type non-linearity is predicted. This non-linearity originates from the Pauli blocking included in the electron-phonon scattering and results in a non-linearly increasing transmission of the film, detectable in experiments.
\end{abstract}
\maketitle
\section{\label{sec:intro} Introduction}
\begin{figure}
    \includegraphics[width=.65\linewidth]{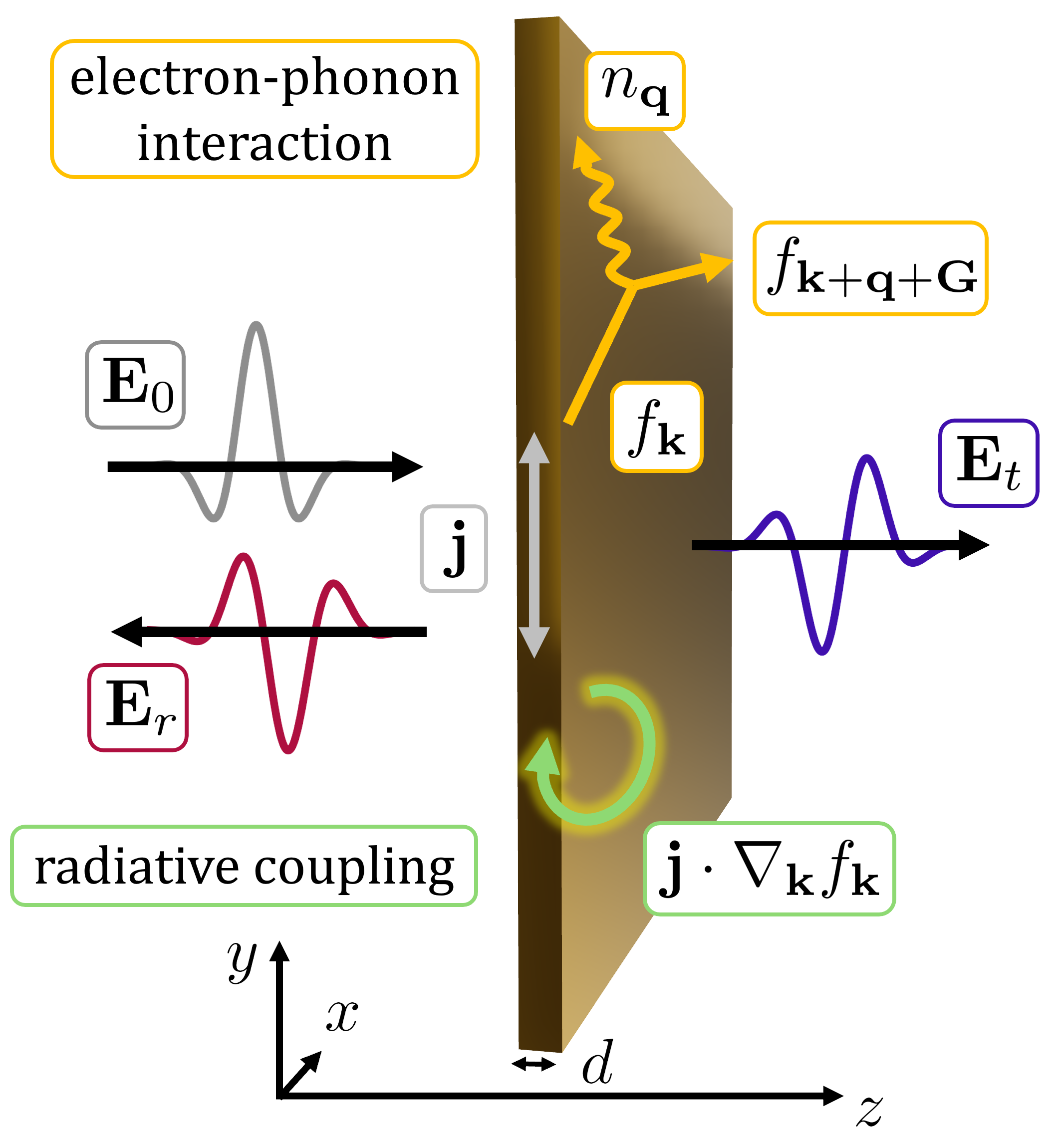} 
    \caption{\label{fig:thin_film} An external light field $\vb{E}_0$ excites a current density $\vb{j}$, determined by a non-equilibrium conduction band electron occupation $f_\mathbf{k}$ in a thin gold film of thickness $d$. Excited electrons are damped via a radiative self-interaction $\vb{j} \cdot \nabla_\mathbf{k} f_\mathbf{k}$, cf.~Eq.~\eqref{eq:thin_film_boltzmann} below, and relax via scattering with phonons $n_\mathbf{q}$ into a state $f_\mathbf{k+q+G}$. The transmitted and reflected fields $\vb{E}_t$ and $\vb{E}_r$ depend on these microscopic processes included in the current density $\vb{j}$.}
\end{figure}

Noble metal plasmonic nanostructures can condense light on the nanoscale and thus enhance light–matter interaction, for example for surface-enhanced Raman scattering, optical waveguides and non-linear optics \cite{choudhury_material_2018, rivera_lightmatter_2020}.
The strong light-matter interaction also enables fundamental time-resolved studies on optical frequencies \cite{dombi_strong-field_2020, wong_far-field_2024}. 
Localized plasmonic excitations are also intensely studied in the context of energy transformation. Many studies have demonstrated that the electron relaxation processes following the plasmon decay can increase the reaction rates of chemical reactions \cite{zhan_plasmon-mediated_2023, aslam_catalytic_2018}. The interaction of light with plasmonic nanostructures is commonly discussed in the linear regime. However, strong exciting fields might lead to non-linear interactions \cite{dombi_strong-field_2020, kim_recent_2022, haas_plasmon_2023, klein_2d_2019}.\\
Firstly formulated in 1900 by Paul Drude, his model provides a classical description of the motion of quasi-free conduction band electrons in noble metals by introducing a phenomenological collision time \cite{drude_zur_1900, maier_plasmonics_2007}. The Drude model has been confirmed by many experimental studies in linear optics \cite{ordal_optical_1987, bennett_colloquium_1965, olmon_optical_2012, weaver_optical_1981, johnson_optical_1972, theye_investigation_1970, blanchard_highresolution_2003}. Strong optical excitations induce non-equilibrium electron distributions beyond thermal electron distributions and require more advanced theories to describe the following relaxation or non-linear optical phenomena \cite{dubi_hot_2019, schirato_ultrafast_2023, khurgin_hot-electron_2023, kauranen_nonlinear_2012}.
\\
Typical non-linear optical properties in noble metals or plasmonic nanostructures can traced back into two origins:
Hot electrons are generated by electronic interband transitions and are energetically in the near-infrared and optical regime. They induce a thermo-modulation of the optical dielectric constant resulting in a non-linear optical response \cite{hache_optical_1988, boyd_third-order_2014, marini_ultrafast_2013, conforti_derivation_2012, khurgin_hot-electron_2023, kauranen_nonlinear_2012}. 
In addition, the spatial confinement of the electron gas in plasmonic nanostructures breaks the translation symmetry at the interfaces between metal and surrounding dielectrics and allows the excitation of high-harmonics or Kerr non-linearities \cite{krasavin_freeelectron_2018, scalora_second-_2010, zeng_classical_2009, rossetti_origin_2024, de_luca_free_2021, cox_analytical_2017, panoiu_nonlinear_2018, fitzgerald_quantum_2016, rodriguez_echarri_nonlinear_2021, christensen_kerr_2015}. 
\\
In this work, we will study the relaxation dynamics of non-equilibrium electrons after optical excitations in thin noble metal films. For the quantitative evaluations, we chose gold as most established plasmonic material and we introduce a third type of optical non-linearity as consequence of the quantum mechanical nature of conduction electrons in a non-equilibrium many-body system:
\\
First, to address the dynamics of non-equilibrium electrons, we solve the kinetic momentum-resolved Boltzmann equation numerically \cite{allen_theory_1987, sivan_theory_2021, del_fatti_nonequilibrium_2000, pietanza_non-equilibrium_2007, rethfeld_ultrafast_2002, mueller_relaxation_2013, seibel_time-resolved_2023, riffe_excitation_2023, ono_ultrafast_2020, sarkar_electronic_2023} within a self-consistent framework with Maxwell's equations for the geometry of thin noble metal films \cite{malic_graphene_2013, stroucken_coherent_1996} displayed in Fig.~\ref{fig:thin_film}. For investigating the relaxation dynamics after optical excitations, we refer to the processes of orientational relaxation, introduced in Ref.~\cite{grumm_ultrafast_2025}, thermalization and cooling, and connect them to detectable optical observables such as transmission, susceptibility or photoluminescence. The processes are distinguished in momentum relaxation (orientational relaxation) which exists independently from the excitation strength and energy relaxation (thermalization and cooling) occurring only if a significant amount of energy is absorbed. So far, to our knowledge, non-equilibrium electron dynamics in noble metals were studied energy-resolved \cite{allen_theory_1987, del_fatti_nonequilibrium_2000, pietanza_non-equilibrium_2007, rethfeld_ultrafast_2002, mueller_relaxation_2013, seibel_time-resolved_2023, riffe_excitation_2023, ono_ultrafast_2020, sarkar_electronic_2023} but not momentum-resolved which excluded the possibility of investigating orientational relaxation.
Focusing on not too strong electron redistribution in comparison to the initial equilibrium, we restrict the theory on electron-phonon interaction which supports all three relaxation processes \cite{mueller_relaxation_2013, grumm_ultrafast_2025} and acts as dominant mechanism for orientational relaxation \cite{lawrence_electron-electron_1973, czycholl_solid_2023} as fundamental process for the dephasing after optical excitations (see Sec.~\ref{sec:IR}): In contrast electron-electron interaction contributes only via umklapp processes to the orientational relaxation \cite{lawrence_electron-electron_1973, czycholl_solid_2023} and its strength decreases much faster than electron-phonon coupling for large momentum transfer \cite{mahan_solid_2000}.
\\
On this footing, many-body non-linearities as they occur in the Pauli blocking of electron-phonon interaction can be analyzed independently of their simultaneous action with electron-electron interaction. We present how polarized excitation fields generate a momentum-polarized non-equilibrium electron gas and reveal a non-linearity in the orientational relaxation dynamics. We predict that this high order dissipative Kerr-type non-linearity is reflected in an increase of the transmission of a thin gold film in the THz regime for strong THz-electric fields of about $1~\mvcm$. 
\\
This paper is structured as follows: We start in Sec.~\ref{sec:micro} and \ref{sec:e_ph_scattering} by introducing our microscopic framework to derive a momentum-resolved Boltzmann equation including electron-phonon scattering via normal and umklapp processes. In Sec.~\ref{sec:maxwell}, we solve the wave equation for the radiation field within a thin film geometry, resulting in self-consistent thin film Maxwell-Boltzmann equations in Sec.~\ref{sec:max_boltz}. Our numerical results will be presented and discussed in Sec.~\ref{sec:results}. Starting with the regime of infrared excitations (Sec.~\ref{sec:IR}), we discuss orientational relaxation, thermalization and cooling for linear and non-linear excitations. Then, in Sec.~\ref{sec:THz}, THz excitations are applied, where the oscillation period of the field, and therefore the time on which the electron gas becomes momentum-polarized, is larger than the typical orientational relaxation time. This reveals a THz field-induced Pauli blocking non-linearity in the transmission spectrum.
Concluding remarks are found in Sec.~\ref{sec:conclusion}.
\section{\label{sec:theo_background} Theoretical Framework}
In this section, we develop the theoretical framework for the description of conduction electrons in noble metals interacting with phonons and the total radiation field after excitation with an external field. To include the re-emission of radiation, we apply a self-consistent formalism of microscopic electron dynamics and macroscopic Maxwell's equations to noble metal films thin compared to the excitation wavelength. A sketch of the used geometry and calculated observables is provided in Fig.~\ref{fig:thin_film}.
\subsection{\label{sec:micro} Microscopic Approach}
In the following, we introduce the used framework for the description of intraband excitations in gold:
In gold, experiments and \textit{ab initio} calculations reveal a band gap of $1.8$~eV between the full-occupied valence \textit{d}-band and half-occupied conduction \textit{sp}-band \cite{johnson_optical_1972, rangel_band_2012, brown_ab_2016}. Therefore, for optical carrier frequencies $\omega_L$ sufficiently below the band gap a description exclusively by intraband transitions in the conduction band is sufficient. The starting point is the single band Hamiltonian \cite{meier_coherent_1994, kira_quantum_1999, rossi_theory_2002, axt_femtosecond_2004, kira_semiconductor_2011}
\begin{align}
    &H = \sum_{\vb{k}} \epsilon_{\vb{k}} a^\dagger_\mathbf{k} a_\mathbf{k} + \frac{ie}{V} \sum_\mathbf{k,q} \vb{E}^{tot}_\mathbf{-q} \cdot a^\dagger_{\vb{k}-\half{q}} \nabla_\mathbf{k} a_{\vb{k}+\half{q}}  \label{eq:hamiltonian}\\
    &+ \sum_{\mathbf{q}} \hbar \omega_{\mathbf{q}} b^\dagger_{\mathbf{q} } b_{\mathbf{q} } + \sum_{\vb{k,q,G}} g_{\mathbf{q}}^\mathbf{G} (b^\dagger_{\mathbf{-q}} + b_{\mathbf{q}}) a^\dagger_\mathbf{k+q+G}a_\mathbf{k} ~. \nonumber
\end{align}
The first term introduces the electronic single particle energy $\epsilon_\mathbf{k}$ with momentum $\vb{k}$ in the conduction band and the corresponding electron annihilation (creation) operators $a^{(\dagger)}_\mathbf{k}$. For simplicity we assume a Sommerfeld free electron model with a parabolic and isotropic dispersion $\epsilon_\mathbf{k} = \frac{1}{2m} \hbar^2|\vb{k}|^2$ with an effective mass $m$ \cite{ashcroft_solid_1976, rethfeld_ultrafast_2002} and a Fermi energy $E_F$. In gold, this assumption is sufficient for heated electron distributions with temperatures up to $T_e<2000$~K \cite{brown_ab_2016, lin_electron-phonon_2008, sundararaman_theoretical_2014}. The interaction between the Fourier components of the total radiation field $\mathbf{E}^{tot}_{\vb{-q}}$, resulting from the external and re-emitted fields at the position of the sample, and the conduction electrons is described by the second term in Eq.~\eqref{eq:hamiltonian} via an acceleration of electrons in direction of the polarization along the conduction band. Here, the total radiation field $\vb{E}_{-\vb{q}}^{tot}$ is determined self-consistently from Maxwell's equations, cf.~Sec.~\ref{sec:maxwell} below, and includes a radiative self-interaction caused by the accelerated electrons. The momentum gradient $-ie\nabla_\mathbf{k}$ is derived for the dipole interaction and corresponds to the intraband dipole operator \cite{gu_relation_2013, foreman_theory_2000, salzwedel_theory_2023} (with the electron charge $q=-e$ and $e>0$). The third term introduces acoustic phonons with bosonic annihilation (creation) operators $b^{(\dagger)}_{\mathbf{q}}$. A linear, isotropic phonon dispersion $\hbar \omega_{\mathbf{q}} = \hbar c_{LA} |\vb{q}|$ with the velocity of sound waves $c_{LA}$ and the phonon momentum $\vb{q}$ is applied. In earlier studies \cite{del_fatti_nonequilibrium_2000}, no significant sensitivity of the electron-phonon scattering processes with respect to phonon dispersion was observed. The coupling of electrons and phonons is given by the last term in Eq.~\eqref{eq:hamiltonian}. In general, in noble metals with mono-atomic basis exist only acoustic phonon modes, which are composed of two transversal (TA) and one longitudinal (LA) branches. Electron-phonon coupling exists for normal processes exclusively for LA phonons and the coupling to TA phonons via umklapp processes is unessential \cite{ziman_electrons_1960, ashcroft_solid_1976, mahan_solid_2000}. In the following, we will consider only LA phonons including their normal and umklapp processes. We apply a screened matrix element \cite{mahan_solid_2000, ashcroft_solid_1976, ziman_electrons_1960} 
\begin{align}
    \vert g_{\mathbf{q} }^{\vb{G}}\vert^2 = \frac{1}{V} \frac{e^2}{2 \varepsilon_0} \frac{\hbar \omega_{\mathbf{q} }}{\vert \vb{q+G} \vert^2 + \kappa^2} \label{eq:def_eph_coupling}
\end{align}
with the Thomas-Fermi screening vector of the Coulomb interaction between electrons and phonons
\begin{align}
    \kappa^2 = \frac{e^2}{\varepsilon_0} \int \mathrm{d}\epsilon_\mathbf{k} ~ D(\epsilon_\mathbf{k}) \frac{\mathrm{d} f_\mathbf{k}^{eq}}{\mathrm{d} \epsilon_\mathbf{k}} ~. \label{eq:screening wavevector}
\end{align}
Here, $D(\epsilon_\mathbf{k})$ denotes the electronic density of states (DOS) and $f_\mathbf{k}^{eq}$ the equilibrium electron distribution \cite{ashcroft_solid_1976, mueller_relaxation_2013}. 
The electron-phonon Hamiltonian in Eq.~\eqref{eq:hamiltonian} includes both normal ($\vb{G}=0$) and umklapp ($\vb{G}\neq0$) processes described by the sum over all reciprocal lattice vectors $\vb{G}$. For phonons described in a Debye model within a reduced zone scheme, this means their momentum $|\vb{q}|$ has to be in the Debye sphere with radius $q_{D} = (6\pi^2 n_0)^{\frac{1}{3}}$ as discussed in App.~\ref{app:umklapp}. 
The parameters used in the Hamiltonian in Eq.~\eqref{eq:hamiltonian} for gold are listed in Tab.~\ref{tab:parameters}. 
\\
The response of the electronic system in Maxwell's equations is determined via the electron density $n$ and the charge current density $\vb{j}$:
\begin{align}
    n (\vb{r}, t) =& -\frac{2}{V} \sum_\mathbf{k} f_\mathbf{k}(\vb{r},t) ~, \nonumber \\
    \vb{j}(\vb{r},t) =& -\frac{2 e}{V} \sum_\mathbf{k} \vb{v_k} f_\mathbf{k}(\vb{r},t) ~. \label{eq:coarse_grained}
\end{align}
Here, $\vb{v_k} = \frac{\nabla_\mathbf{k} \epsilon_\mathbf{k}}{\hbar}$ is the electron velocity determined by the band structure. In general, the sum includes also the spin. Since the initial optical excitation is not spin-sensitive, we study here spin-degenerated electrons with respect to electron-phonon scattering and the spin-sum results in the factor $2$ in Eq.~\eqref{eq:coarse_grained} \cite{mueller_relaxation_2013, czycholl_solid_2023}. 
Here, the Wigner function $f_\mathbf{k}(\vb{r},t)$ as quantum-mechanical electron phase-space density \cite{wigner_quantum_1932, kira_semiconductor_2011} is introduced as microscopic observable
\begin{equation}
    f_\mathbf{k}(\vb{r},t) = \sum_\mathbf{q} e^{i\vb{q}\cdot\vb{r}} \langle a^\dagger_{\vb{k}-\half{q}} a_{\vb{k}+\half{q}} \rangle(t) = \sum_\mathbf{q} e^{i\vb{q}\cdot\vb{r}} \Tilde{f}_{\vb{k}-\half{q}, \vb{k}+\half{q}}  ~.
\end{equation}
The time-dependent expectation value of the Wigner function is calculated in the Heisenberg equation of motion framework with the Hamiltonian from Eq.~\eqref{eq:hamiltonian} and yields the kinetic Boltzmann equation similar to Refs.~\cite{hess_maxwell-bloch_1996, meier_coherent_1994, del_fatti_nonequilibrium_2000, pietanza_non-equilibrium_2007, mueller_relaxation_2013, riffe_excitation_2023, salzwedel_theory_2023, rethfeld_ultrafast_2002}
\begin{equation}
    \partial_t f_\mathbf{k}(t) = - \vb{v_k} \cdot \nabla_\mathbf{r} f_\mathbf{k}(t) + \frac{e}{\hbar} \vb{E}^{tot}(\vb{r},t) \cdot \nabla_\mathbf{k} f_\mathbf{k}(t) + \Dot{f}_\mathbf{k}(t)\vert_{scatt} ~. \label{eq:boltzmann}
\end{equation}
The first term results from the electronic single-particle Hamiltonian in a gradient expansion as outlined in Ref.~\cite{hess_maxwell-bloch_1996}. The intraband light-matter coupling yields the second term describing electron acceleration in the conduction band in response to the distinct polarization direction of the total radiation field. The last term contains dissipation due to electron-phonon scattering, i.e.~relaxation via momentum and energy exchange.
\subsection{\label{sec:e_ph_scattering} Electron-Phonon Scattering} 
Applying the electron-phonon Hamiltonian in Eq.~\eqref{eq:hamiltonian}, the dissipative contribution of the Boltzmann equation is derived in a quantum kinetic treatment in second order Born approximation \cite{haug_quantum_2008}
\begin{widetext}
    \begin{align}
        \partial_t f_\mathbf{k}(t)\vert_{scatt} = -\frac{i}{\hbar} &\sum_{\mathbf{q}, \vb{G},\pm} |g_{\mathbf{q} }^{\vb{G}}|^2 \times \label{eq:eom_eph} \\
        &\times\Big(\pm \frac{i}{\hbar} \int_{-\infty}^t \mathrm{d}t'~ e^{-\frac{i}{\hbar} \big(\epsilon_{\vb{k}_1} - \epsilon_{\vb{k}_2} + \hbar \omega_{\mathbf{q}} \big) (t-t') } \big\{ (1+n_{\mathbf{q} }) f_{\vb{k}_2}(t') - n_{\mathbf{q} } f_{\vb{k}_1}(t') - f_{\vb{k}_1}(t') f_{\vb{k}_2}(t') \big\} \nonumber \\
        &~~~~~\pm \frac{i}{\hbar} \int_{-\infty}^t \mathrm{d}t'~ e^{-\frac{i}{\hbar} \big(\epsilon_{\vb{k}_1} - \epsilon_{\vb{k}_2} - \hbar \omega_{\mathbf{q}} \big) (t-t') } \big\{ n_{\mathbf{q} } f_{\vb{k}_2}(t') - (1+n_{\mathbf{q} })f_{\vb{k}_1}(t') + f_{\vb{k}_1}(t') f_{\vb{k}_2}(t') \big\} \Big)  \nonumber
    \end{align}
\end{widetext}
with the abbreviations $\vb{k}_1 = \vb{k}-\half{q+G}\pm\half{q+G}$ and $\vb{k}_2 = \vb{k}+\half{q+G}\pm\half{q+G}$. A detailed derivation is discussed in App.~\ref{app:phonon}. In Eq.~\eqref{eq:eom_eph}, the electronic occupations are integrated over all past times $t' < t$, which formally includes memory effects and allows violation of the conservation of energy in terms of the uncertainty principle \cite{haug_quantum_2008}. In the Markov approximation \cite{rossi_theory_2002, haug_quantum_2008, indik_role_1996}, such phenomena are neglected, Eq.~\eqref{eq:eom_eph} becomes local in time, and the electron-phonon scattering terms yield the well-known semi-classical scattering terms in the Boltzmann equation \cite{allen_theory_1987, del_fatti_nonequilibrium_2000, rethfeld_ultrafast_2002, mueller_relaxation_2013, pietanza_non-equilibrium_2007, ono_ultrafast_2020, un_electronic-based_2023, riffe_excitation_2023, mahan_solid_2000}
\begin{equation}
    \partial_t f_\mathbf{k}(t)\vert_{scatt} = \Gamma^{in}_\mathbf{k}(t) (1-f_\mathbf{k}(t)) - \Gamma^{out}_\mathbf{k}(t) f_\mathbf{k}(t) ~, \label{eq:scattering_eq}
\end{equation}
where the scattering rates 
\begin{align}
    \Gamma^{in}_\mathbf{k} =& \frac{2\pi}{\hbar} \sum_{\vb{q}, \vb{G}, \pm} |g_{\mathbf{q} }^{\vb{G}}|^2 \delta(\epsilon_\mathbf{k+q+G}-\epsilon_\mathbf{k} \pm \hbar \omega_{\mathbf{q}}) \times \nonumber\\
    &\times \Big( \frac{1}{2} \mp \frac{1}{2} + n_{\mathbf{q} } \Big) f_\mathbf{k+q+G} ~, \\
    \Gamma^{out}_\mathbf{k} =& \frac{2\pi}{\hbar} \sum_{\vb{q}, \vb{G}, \pm} |g_{\mathbf{q} }^{\vb{G}}|^2 \delta(\epsilon_\mathbf{k+q+G}-\epsilon_\mathbf{k} \pm \hbar \omega_{\mathbf{q}}) \times \nonumber\\
    &\times \Big( \frac{1}{2} \pm \frac{1}{2} + n_{\mathbf{q} } \Big) (1-f_\mathbf{k+q+G}) ~
\end{align}
are time-dependent in terms of $f_\mathbf{k+q+G}(t)$. All terms in the scattering equation depend on the occupation of the initial and final electronic state and on the phonon distribution with the corresponding transferred quasi-momentum $\vb{q}+\vb{G}$ as schematically depicted in Fig.~\ref{fig:thin_film}. To restrict the description to a reduced zone scheme, umklapp processes are considered so that the initial and final electron momenta $\vb{k}$ and $\vb{k+q+G}$ as well as the phonon momentum $\vb{q}$ are in the first Brillouin zone (whereas $\vb{q+G}$ itself is not restricted to the first Brillouin zone) \cite{gross_many-particle_1986}. This conditions restrict the set of permitted reciprocal lattice vectors $\vb{G}$ for electron-phonon scattering as discussed in App.~\ref{app:umklapp}.
\\
As the theory is formulated in a Markovian limit, the rates conserve energy during scattering events represented by the $\delta$-distributions and are described by Fermi's golden rule. In total, there are four terms that contain combinations of phonon absorption and emission together with electronic intraband transitions from a state $\vb{k}$ to $\vb{k+q+G}$ and vice versa. The factors $f_\mathbf{k+q+G}(1-f_\mathbf{k})$ and $(1-f_\mathbf{k+q+G})f_\mathbf{k}$ are characteristic for fermions and describe the Pauli blocking. Note that these terms introduce an intrinsic non-linearity to the electronic occupations $f_\mathbf{k}$. The non-linear Pauli blocking terms arise from a phonon-mediated electronic mean field in a Hartree-Fock-like correlation expansion (cf.~quadratic terms in $f_\mathbf{k}$ in Eq.~\eqref{eq:eom_eph}) \cite{fricke_transport_1996} as result of many-body effects. 
\\
The phonons are treated as a thermal bath described via a static Bose-Einstein distribution 
\begin{align}
    n_{\mathbf{q} } = \frac{1}{\exp{\frac{\hbar \omega_{\mathbf{q} }}{k_B T_p}}-1} 
\end{align}
at a constant lattice temperature $T_p$. To validate this assumption, we have estimated the dynamics of the phonon temperature by a two-temperature model with typical parameters for gold \cite{hartland_coherent_2002}. For electronic temperatures $T_e \ll T_F$ (Fermi temperature $T_F)$, the heat capacity of the electrons is much smaller than the phononic one and changes in the phonon temperature is negligible \cite{brown_ab_2016, giri_experimental_2015, hartland_optical_2011, lin_electron-phonon_2008}.
\subsection{\label{sec:maxwell} Self-Consistent Radiation Field}
In many experiments the electronic and optical properties of noble metals are studied in thin films of thickness $d$ typically in the range of up to $50$~nm \cite{johnson_optical_1972, olmon_optical_2012}. We restrict to these limit for the film thickness which allow a simple solution of Maxwell's equations in a plane wave geometry. The corresponding geometry is depicted in Fig.~\ref{fig:thin_film}.
The goal of this section is a formulation of a self-consistent solution of the macroscopic Maxwell's equations and the microscopic material dynamics to calculate the total radiation field $\vb{E}^{tot}$ occurring in Eq.~\eqref{eq:boltzmann}. 
\\
To connect the microscopic dynamics in Sec.~\ref{sec:micro} and \ref{sec:e_ph_scattering} to optical observables, we calculate the total radiation field from the wave equation as provided by Ref.~\cite{malic_graphene_2013, stroucken_coherent_1996}
\begin{equation}
    \Big( \Delta - \frac{1}{c^2} \partial_t^2 \Big) \vb{E}^{tot} = \mu_0 \partial_t \vb{j}_s ~. \label{eq:wave_eq}
\end{equation}
The current density $\vb{j}_s$ acts here as general source term. Later in Eq.~\eqref{eq:source_current}, we apply a current density which includes intraband currents as defined in Eq.~\eqref{eq:coarse_grained} as well as an effective dielectric background screening. In the wave equation, $c = \frac{c_0}{n_{out}}$ is the speed of light in the surrounding medium with refractive index $n_{out}$.  
As homogeneous solution of the wave equation we assume an external irradiated in $y$-direction polarized field $\vb{E}_0 = E_0 \vb{e_y}$ as plane wave propagating in $z$-direction, cf.~Fig.~\ref{fig:thin_film}. The inhomogeneous solution is obtained from a Green's function approach in Ref.~\cite{malic_graphene_2013, stroucken_coherent_1996}
\begin{align}
    \vb{E}^{tot} (\vb{r},t) =& \vb{E}_0 \Big(\vb{r}, t-\frac{r}{c} \Big) \nonumber \\
    &- \frac{\mu_0}{4\pi} \partial_t \int \mathrm{d}^3r'~ \frac{\vb{j}_s \big(\vb{r}', t \pm \frac{\vert \vb{r}-\vb{r}'\vert}{c} \big)}{\vert \vb{r}-\vb{r}' \vert} ~, \label{eq:formal_solution}
\end{align}
where the sign of $\pm$ differs between the reflected and transmitted wave. A detailed discussion is provided in App.~\ref{app:induced_field} and leads to the induced field re-emitted by the current density at the position $z=0$
\begin{align}
    \vb{E}_{ind}(t) =  - \frac{\mu_0 c_0 d}{2n_{out}} \vb{j}_s(t) ~. \label{eq:induced_field}
\end{align}
The forward propagating induced field $\vb{E}_{ind}$ together with the irradiated field $\vb{E}_0$ yields the transmitted field $\vb{E}_t$ trough the thin film of thickness $d$, as depicted in Fig.~\ref{fig:thin_film}
\begin{equation}
    \vb{E}_t (z,t) = \vb{E}_0 \Big(t-\frac{z}{c}\Big) - \frac{\mu_0 c_0 d}{2n_{out}} \vb{j}_s\Big(t-\frac{z}{c}\Big) ~. \label{eq:transmitted_field}
\end{equation}
The reflected field is completely determined by the backwards propagating induced field via
\begin{equation}
    \vb{E}_r(z,t) = - \frac{\mu_0 c_0 d}{2n_{out}} \vb{j}_s\Big(t+\frac{z}{c}\Big)  ~. \label{eq:reflected_field}
\end{equation}
The total radiation field $\vb{E}^{tot}$ in the film equals the transmitted field $\vb{E}_t$ at the position of the sample, i.e. Eq.~\eqref{eq:transmitted_field} for $z=0$. This way, Eq.~\eqref{eq:transmitted_field} and Eq.~(\ref{eq:coarse_grained}, \ref{eq:boltzmann}) determine the full self-consistent dynamics.
\\
One benchmark for our approach is the linear response which need to reproduce linear optical spectra of gold films: In this limit, we apply the Fourier transformed current density $\hat{j}_s(\omega) = -i\omega \varepsilon_0 \chi(\omega) \hat{E}^{tot}(\omega)$ with an isotropic material susceptibility $\chi(\omega)$. The transmission $T(\omega)$ and reflection $R(\omega)$ coefficients read
\begin{align}
    T(\omega) = \Big\vert \frac{\hat{\vb{E}}_t(\omega)}{\hat{\vb{E}}_0(\omega)} \Big\vert^2 = \frac{1}{\vert 1- i\frac{\omega d}{2 c_0 n_{out}} \chi(\omega) \vert ^2} \label{eq:T} ~, \\
    R(\omega) =  \Big\vert \frac{\hat{\vb{E}}_r(\omega)}{\hat{\vb{E}}_0(\omega)} \Big\vert^2 = \frac{ \frac{\omega^2 d^2}{4 c_0^2 n_{out}^2} \vert \chi(\omega) \vert^2}{\vert 1- i\frac{\omega d}{2 c_0 n_{out}} \chi(\omega) \vert ^2} ~. \label{eq:R}
\end{align}
These formulae derived from Eq.~(\ref{eq:transmitted_field}, \ref{eq:reflected_field}) are also known in the literature \cite{lanzani_probing_2005, nuss_terahertz_1991, glover_conductivity_1957}. In the limit of a thin film ($\frac{2\omega d n_{out}}{\pi c_0} \ll 1$) embedded in a surrounding medium with refractive index $n_{out}$, the well-known Fresnel formulae for a homogeneous dielectric film \cite{born_principles_2019} become identical with Eq.~(\ref{eq:T},\ref{eq:R}).
\\
We briefly summarize the equations which need to be solved self-consistently in the next section.
\subsection{\label{sec:max_boltz} Self-Consistent Maxwell-Boltzmann Equation}
The total radiation field $\vb{E}^{tot}$ is determined by the sum of the external field $\vb{E}_0$ and the re-emitted field $\vb{E}_{ind}$ in Eq.~\eqref{eq:induced_field} acting at the position of the thin noble metal film at $z \approx 0$. The current density as source term in the wave equation \eqref{eq:wave_eq} is assumed to be 
\begin{align}
    \vb{j}_s(t) = \vb{j}(t) + \vb{j}_b(t) ~. \label{eq:source_current}
\end{align} 
Here, $\vb{j}(t)$ is the current generated by the intraband dynamics of the conduction electrons defined in Eq.~\eqref{eq:coarse_grained}. Motivated by experimental observations \cite{shahbazyan_plasmonics_2013, vial_improved_2005}, we include screening effects due to the dielectric background, e.g.~core level electrons or interband transitions via a current $\vb{j}_b$ approximated by an effective constant $\varepsilon_b$. Thus, a background polarization $\vb{P}_b$ is introduced with $\vb{j}_b = \Dot{\vb{P}}_b = \varepsilon_0 \Tilde{\varepsilon}_b \Dot{\vb{E}}^{tot}(t)$, where $\Tilde{\varepsilon}_b = \varepsilon_b - \varepsilon_{out}$ and $\varepsilon_{out}=n_{out}^2$. This yields a total radiation field
\begin{align}
    &\vb{E}^{tot}(t) = \vb{E}_0(t) \label{eq:total_radiation_field} \\
    &- \frac{\mu_0 c_0 d}{2 n_{out}} \Big( -\frac{2e}{V} \sum_{\vb{k}'} \vb{v}_{\vb{k}'} f_{\vb{k}'}(t) + \varepsilon_0 \Tilde{\varepsilon}_b \Dot{\vb{E}}^{tot}(t) \Big) ~,\nonumber 
\end{align}
where the time derivative on the right hand side is solved during the numerical time integration via a finite difference method.
\\
The total radiation field $\vb{E}^{tot}$, applied to Eq.~\eqref{eq:boltzmann}, results an integro-differential non-linear Boltzmann equation for the electron dynamics in thin films
\begin{align}
    \partial_t f_\mathbf{k}(t) =& \frac{e}{\hbar} \vb{E}^{tot}(t) \cdot \nabla_\mathbf{k} f_\mathbf{k}(t) \nonumber \\
    &+ \Gamma^{in}_\mathbf{k}(t) (1-f_\mathbf{k}(t)) - \Gamma^{out}_\mathbf{k}(t) f_\mathbf{k}(t)  ~. \label{eq:thin_film_boltzmann}
\end{align}
To derive this equation, the spatial gradient parallel to the film in Eq.~\eqref{eq:boltzmann} was neglected, since a perpendicular incidence of the wave on the plane of the film and a thickness small compared to the wavelength of the external field are assumed.
Eq.~(\ref{eq:total_radiation_field}, \ref{eq:thin_film_boltzmann}) represent a self-consistent solution of macroscopic Maxwell's dynamics of the electric fields and microscopic electron dynamics in a plane wave geometry. The term containing the induced field creates a self-interaction of the electron gas leading to a radiative damping of electron dynamics depicted in Fig.~\ref{fig:thin_film}. However, the radiative damping of the macroscopic current density derived in Eq.~\eqref{eq:derivation_gamma_rad} depends only on the total electron density which is conserved in a spatially homogeneous system. Thus, as validated by our numerical results in Sec.~\ref{sec:IR}, the self-interaction term, bracket in Eq.~\eqref{eq:total_radiation_field}, does not contribute to the non-linearities in the optical response. In contrast, electron-phonon scattering (Pauli blocking in Eq.~\eqref{eq:scattering_eq}) introduces optically detectable many-body non-linearities to the microscopic electron dynamics as we will discuss in the following. 
\section{\label{sec:results} Results and Discussion} 
\begin{figure}
    \centering
    \includegraphics{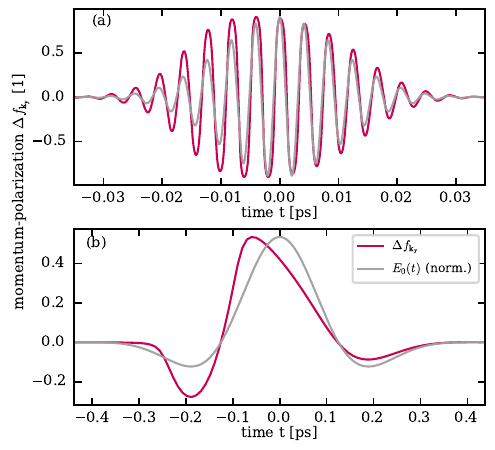}
    \caption{In the infrared regime ($\hbar \omega_L=1$~eV) in (a), the dynamics of the momentum-polarization $\Delta f_\mathbf{k_y}(t)$ is in phase with the incident field $E_0(t)$. In (b) for THz excitations ($\hbar \omega_L=8$~meV, $\nu_L=2$~THz), where the orientational relaxation acts faster than the oscillation of the field, this phase relation is broken. In both cases an external field strength of $E_0=5~\mvcm$ is applied.}
    \label{fig:distribution}
\end{figure}

In Ref.~\cite{grumm_ultrafast_2025}, we discuss the relaxation of an optically excited non-equilibrium electron gas via electron-phonon interaction in a bulk material in terms of three processes: Orientational relaxation, thermalization, and cooling. In a first step, orientational relaxation is the relaxation of the direction of momentum and can only be observed in a study of electron dynamics in the three dimensional momentum space. Orientational relaxation emerges from the momentum-polarization of the electron gas induced by the polarization direction of the incident field. In contrast, the following thermalization and cooling, both representing energy relaxation processes, depend only on the electron energy and can be reduced to an energy-resolved dynamics only. In addition, radiative damping occurs in a thin film geometry due to the self-interaction of the electrons via the radiation field here. In typical thin films with thicknesses of several nm to a few tens of nm \cite{johnson_optical_1972, olmon_optical_2012}, the de Broglie wavelength of the electrons is small compared to the thickness, size quantization effects for electrons can be neglected \cite{rodriguez_echarri_quantum_2019}, and the used bulk model is valid.
\\
Importantly, to disentangle orientational and energy relaxation, the field strength acting on the electrons $E^{tot}$ determines whether the excitation is linear or non-linear and therefore whether energy relaxation occurs. To distinguish between the linear (orientational relaxation) and non-linear (orientational and energy relaxation) regimes, the energy absorbed by the electrons
\begin{align}
    \epsilon_{abs} = \Big| E_F - \frac{\hbar^2}{2m} (k_F - k_{abs})^2 \Big| 
\end{align}
is estimated via the optically excited momentum $k_{abs} = -\frac{e}{\hbar} \int_{-\infty}^{\infty} \mathrm{d}t ~|E^{tot}(t)|$ as derived from the classical Lorentz force \cite{ziman_electrons_1960}. We define an excitation as non-linear if the corresponding absorbed energy is at least of the same order of magnitude as the thermal broadening of the equilibrium electron distribution ($\epsilon_{abs} \gtrsim k_B T_{eq}$). Since these excitations induce strong non-equilibrium conditions in the electron gas, the Pauli blocking terms in Eq.~\eqref{eq:scattering_eq} turn on and influence the relaxation dynamics. In contrast, the linear regime is valid for $\epsilon_{abs} \ll k_B T_{eq}$. Since a negligible amount of energy is absorbed by the electron gas in the linear regime, only orientational relaxation (momentum relaxation) appears in linear optical observables, cf.~Fig.~\ref{fig:thin_film}, but not thermalization and cooling as energy relaxation.
\\
In the following two limits are studied, distinguished by the frequency of the incident light field $\omega_L$ relative to the typical orientational relaxation time $\tau_{or}$ of tens of fs \cite{grumm_ultrafast_2025}: In Sec.~\ref{sec:IR}, infrared excitations are discussed with $\omega_L^2 \gg \tau_{or}^{-2}$, where in the THz regime in Sec.~\ref{sec:THz} $\omega_L^2 \ll \tau_{or}^{-2}$ is valid.
\\
In a microscopic picture, this distinction is motivated by the time on which the momentum-polarization of the electron gas is modulated by the incident field with frequency $\omega_L$. Each half oscillation period of the linearly polarized incident field, its polarization direction switches in sign. This switch in polarization is also imprinted on the dynamics of the momentum-polarized electron gas, so that the strength of momentum-polarization is determined by the frequency of the incident light field and not on the temporal pulse duration. To characterize this dynamics, we define the momentum-polarization between two momentum-states located at the Fermi edge $k_F$ in direction of the incident field polarization $\vb{e_y}$ by $\Delta f_\mathbf{k_y} = f_{-k_F\vb{e_y}} - f_{k_F\vb{e_y}}$. Fig.~\ref{fig:distribution} depicts this momentum-polarization, excited by the incident field and competing with orientational relaxation as calculated numerically from Eq.~\eqref{eq:thin_film_boltzmann}. For infrared excitation in Fig.~\ref{fig:distribution}a, the momentum-polarization follows the dynamics of the incident field with the same phase, since the orientational relaxation is slower than the oscillation frequency of the field. 
In contrast, for THz excitation in Fig.~\ref{fig:distribution}b, the orientational relaxation occurs faster than the oscillation period of the incident field and results in a decay of the momentum-polarization during each oscillation. This generates the asymmetric dynamics depicted in Fig.~\ref{fig:distribution}b and prevents a constant phase relation.
\\
Furthermore, we will show below that orientational relaxation is the microscopic mechanism of the phenomenologically introduced damping $\gamma_D$ in the Drude model for the optical response of noble metals $\chi_D$ (cf.~App.~\ref{app:drude}). There, the convenient approximation in the near-infrared and optical regime is $\omega^2 \gg \gamma_D^2$ or vice versa in the THz regime, which leads to fundamental different frequency dependencies in both regimes
\begin{align}
    \chi_D(\omega) \approx 
    \begin{cases}
        \Tilde{\varepsilon}_b - \frac{\omega_p^2}{\omega^2} + i\frac{\omega_p^2 \gamma_D}{\omega^3} ,~\text{IR}~(\omega^2 \gg \gamma_D^2) \\ 
        \Tilde{\varepsilon}_b - \frac{\omega_p^2}{\gamma_D^2} + i \frac{\omega_p^2}{\omega \gamma_D} ,~\text{THz}~(\omega^2 \ll \gamma_D^2) ~, \label{eq:drude_limits}
    \end{cases} 
\end{align}
where $\omega_p$ is the plasma frequency. This shows that the real part of the susceptibility dominates in the infrared range, while the imaginary part increases in relevance at THz frequencies. Thus, optical observables are far more sensitive to the orientational relaxation  rate in the THz regime than in the infrared regime. This motivates to investigate in the following linear and non-linear orientational relaxation in cases where the built-up of the momentum-polarization is short compared to the typical orientational relaxation time (IR regime, Sec.~\ref{sec:IR}) or where the built-up is much slower than the orientational relaxation (THz regime, Sec.~\ref{sec:THz}).
\\
In this way, we discuss in the following the numerical solution of the thin film Boltzmann equation \eqref{eq:thin_film_boltzmann} in these regime with parameters for gold listed in Tab.~\ref{tab:parameters}.
\subsection{\label{sec:IR} Infrared Excitations}
The aim of this section is to analyze the simultaneous action of all the processes introduced above after excitation with ultrashort infrared pulses of $24$~fs duration FWHM. Similar ultrashort excitations are commonly used in experiments \cite{suemoto_observation_2019, suemoto_relaxation_2021, sun_femtosecond_1993, sun_femtosecond-tunable_1994} and other theoretical studies \cite{mueller_relaxation_2013, brown_experimental_2017, del_fatti_nonequilibrium_2000} to trace energy relaxation, since the excitation time is short compared to the expected energy relaxation times \cite{grumm_ultrafast_2025}. In this way, the processes of orientational relaxation and energy relaxation (thermalization and cooling) can be discussed in a sequential temporal order. To connect this time hierarchy to experiments, we study the time and spectral dynamics of the transmitted radiation field, the optical susceptibility, and the photoluminescence and show how the relaxation processes as characteristic quantities of the electron gas are related to these optical observables. 
\\
In the following, a gold film with thickness $d=20$~nm and an incident electric field with carrier frequency $\hbar \omega_L = 1.0$~eV (wave length $\lambda_L = 1.24~\mu\text{m}$) is considered. The excitation energy of $1.0$~eV allows the restriction to an intraband description. 
For this, we observe significant non-linear effects for an external field strength of $E_0 = 20~\mvcm$ comparable to other theoretical studies in this frequency regime \cite{mueller_relaxation_2013, pietanza_non-equilibrium_2007, rethfeld_ultrafast_2002, del_fatti_nonequilibrium_2000, brown_experimental_2017} where thermalization and cooling are discussed. In the linear regime, we assume here exemplarily an external field strength of $E_0/100$. Applying these parameters for the external field, in Sec.~\ref{sec:IR_linear} linear and in Sec.~\ref{sec:IR_nonlinear} non-linear effects are discussed.
\subsubsection{\label{sec:IR_linear} Linear Excitation Regime}
\begin{figure}
    \centering
    \includegraphics{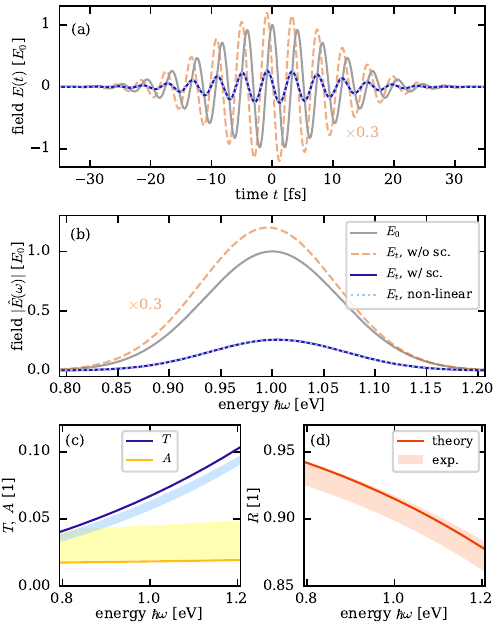}
    \caption{The incident field $E_0$ and transmitted field $E_t$ calculated with and without self-consistency (sc.) in the thin film Boltzmann equation \eqref{eq:thin_film_boltzmann} are shown in (a) time- and (b) frequency-domain. The (normalized) transmitted field after nonlinear excitations (dotted lines, see discussion in section ~\ref{sec:IR_nonlinear}) displays no differences to the linear regime. In (c,d) the calculated transmission, reflection and absorption coefficients ($T,~R,~A$, solid lines) are compared with typical experimental data (shaded areas) \cite{johnson_optical_1972, weaver_optical_1981, born_principles_2019}.}
    \label{fig:IR_spectrum}
\end{figure}

The external field pulse $E_0/100$ (cf.~discussion above) propagates trough the thin gold film, excites the electron gas, self-interacts via this excitation, and gets transmitted. For a detailed understanding of this, Fig.~\ref{fig:IR_spectrum}a,b compare the incident field $E_0$ (grey) and the transmitted field $E_t$ (solid blue line) calculated self-consistent from Eq.~\eqref{eq:transmitted_field} in time and frequency domain. Fig.~\ref{fig:IR_spectrum}c,d visualize the corresponding transmission $T$ and reflection coefficients $R$ calculated with Eq.~(\ref{eq:T}, \ref{eq:R}) as well as the absorption $A=1-R-T$ (solid lines).
\\
The reduction in amplitude of the transmitted field is caused by the dominating reflection of metals in the infrared regime  (cf.~Fig.~\ref{fig:IR_spectrum}d) and by the internal absorption of the gold film via dissipative electron-phonon scattering (cf.~the absorption in Fig.~\ref{fig:IR_spectrum}c). The relevance of the inclusion of self-consistency of Maxwell's and Boltzmann equations is visualized by the transmitted field calculated without self-consistency (dashed line in Fig.~\ref{fig:IR_spectrum}a,b). For this, we have neglected the self-interaction term in the thin film Boltzmann equation \eqref{eq:thin_film_boltzmann}, i.e.~both terms in the bracket in Eq.~\eqref{eq:total_radiation_field}, which leads to non-physical results, namely a transmitted field that is larger than the incident field. Therefore, in order to calculate realistic observables, a self-consistent treatment of Maxwell's equations and electron dynamics is necessary.
\\
In addition, in Fig.~\ref{fig:IR_spectrum}a, a phase shift and weak frequency modulation in the self-consistent transmitted field is observed. This frequency modulation leads to a weak blue-shift of the transmitted field spectrum in comparison to the external field spectrum in Fig.~\ref{fig:IR_spectrum}b. For the parameters selected in the infrared regime, this shift is less than $1\%$ of the frequency of the external field $\omega_L$ but is not observed in the transmitted field without self-consistency. The blue-shift is in agreement with calculations based on the Fresnel formulae for the transmitted field trough a homogeneous thin film \cite{born_principles_2019}. 
Furthermore, our numerical results for transmission, reflection and absorption, calculated in Fig.~\ref{fig:IR_spectrum}c,d via the thin film approach from Sec.~\ref{sec:maxwell}, are in good agreement with typical experimental data (shaded areas). For a comparison of theory and experiment, we used experimental data for the optical dielectric function of gold \cite{johnson_optical_1972, olmon_optical_2012, weaver_optical_1981} together with the Fresnel formulae \cite{born_principles_2019}, which include internal Fabry-Pérot interferences in the film. For better illustration, we used a Drude fit, Eq.~\eqref{eq:chi_drude}, of the experimental data, which shows excellent agreement with the original data. The deviations between theory and experiment can be explained by the fact, that our numerical approach neglects internal interferences in the limit $\frac{2 \omega d n_{out}}{\pi c_0} \ll 1$.

\begin{figure}
    \centering
    \includegraphics{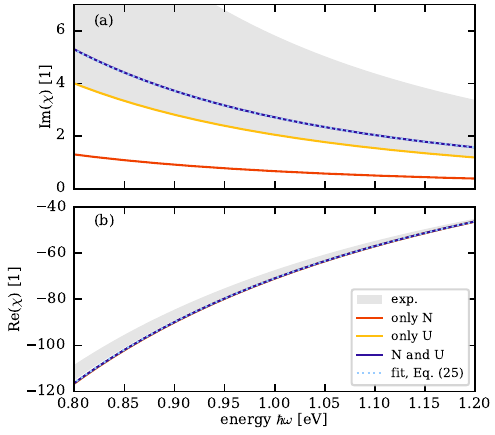}
    \caption{Comparison of the (a) imaginary and (b) real part of the optical susceptibility $\chi$ for normal (N) and umklapp (U) processes in the electron-phonon interaction. Including both processes results in a susceptibility in agreement with typical experimental data (shaded areas, data adapted from Ref.~\cite{weaver_optical_1981, johnson_optical_1972}).}
    \label{fig:susceptibility}
\end{figure}

Due to the optical excitation, the electron gas is momentum-polarized in direction of the incident field polarization and an anisotropic momentum distribution can be observed over the whole Brillouin zone \cite{grumm_ultrafast_2025}. In the following, orientational relaxation sets in and the momentum is re-distributed by electron-phonon scattering until an isotropic momentum distribution is achieved. This process can already be observed for linear excitations independently of the absorbed energy: The finite current density $\vb{j}$ as defined in Eq.~\eqref{eq:coarse_grained} results from the momentum-polarized electron gas. This intraband current density is generated only by anisotropically distributed electrons \cite{hess_maxwell-bloch_1996-1} and decays within the \textit{orientational relaxation time} $\tau_{or}$ \cite{grumm_ultrafast_2025}. In the limit of linear optics, the Fourier transformed total current density in the thin film $\hat{\vb{j}}_s(\omega)$, Eq.~\eqref{eq:source_current}, is related by Ohm's law with the total radiation field $\hat{\vb{E}}^{tot}$ at the position of the sample
\begin{align}
    \hat{\vb{j}}_s(\omega) = -i\omega \varepsilon_0 \chi(\omega) \hat{\vb{E}}^{tot}(\omega) ~. \label{eq:susceptibility}
\end{align}
As shown in App.~\ref{app:drude}, the susceptibility $\chi(\omega)$ is connected in the linear regime with a Drude model introducing an effective electron-phonon momentum relaxation rate also obtained from the well-known Bloch-Grüneisen relation \cite{smith_frequency_1982, mahan_solid_2000, czycholl_solid_2023}. Thus, the imaginary part of the susceptibility in Eq.~\eqref{eq:susceptibility} encodes the orientational relaxation time $\tau_{or}$. Fig.~\ref{fig:susceptibility} illustrates the susceptibility derived from the numerical solution of the thin film Boltzmann equation \eqref{eq:thin_film_boltzmann} including normal and umklapp processes (solid blue lines). Both, real and imaginary part agree with the fitted Drude susceptibility $\chi_D$ from Eq.~\eqref{eq:chi_drude} (dotted lines). 
\\
To determine the orientational relaxation time from the susceptibility, we adapt a method from Ref.~\cite{johnson_optical_1972}: Analyzing $\omega \Im{\chi}$ as function of $\frac{1}{\omega^2}$ results in the limit $\tau_{or}^{-2} \ll \omega^2$, as it is valid in the infrared regime, in a linear relation for a Drude-like material response
\begin{align}
    \omega \Im(\chi_D) = \omega_p^2 \frac{\tau_{or}^{-1}}{\omega^2 + \tau_{or}^{-2}} \approx \omega_p^2 \frac{\tau_{or}^{-1}}{\omega^2} ~. \label{eq:im_drude_chi}
\end{align}
With this, we determine $\tau_{or} = 19.6$~fs.
By linearizing the scattering equation \eqref{eq:scattering_eq}, a semi-analytical expression for the momentum orientational relaxation time is derived in App.~\ref{app:drude} in agreement with the numerical results.
\\
In addition, Fig.~\ref{fig:susceptibility} compares the susceptibility in the context of normal (N) and umklapp (U) processes in the electron-phonon interaction. Whilst the real part is, as expected in the infrared regime (cf.~Eq.~\eqref{eq:drude_limits}), unaffected by changes in the electron-phonon interaction and all curves superimpose, there occur significant differences in the imaginary part: Considering for the numerical solution only normal processes (red) yields $\tau_{or}^N = 80.1$~fs, whereas only umklapp processes (yellow) result in $\tau_{or}^U = 26.0$~fs. Respecting both processes (blue) results in agreement with the Matthiessen's rule \cite{ziman_electrons_1960} in $\tau_{or} = \big(\frac{1}{\tau_{or}^N} + \frac{1}{\tau_{or}^U}\big)^{-1} = 19.6$~fs. The dominance of umklapp processes by a factor of three against normal processes to the orientational relaxation is documented in literature for analytical estimates \cite{lawrence_umklapp_1972} and is easy to understand by considering how normal and umklapp processes act in a momentum-polarized electron gas:
\\
As discussed in the context of Fig.~1 in Ref.~\cite{grumm_ultrafast_2025}, the momentum-polarization of the anisotropic electron distribution is maximized for momenta parallel to the polarization direction $\vb{e_y}$ of the incident field $\vb{E}_0$ at opposite points on the Fermi surface, i.e.~for electrons with momenta $+k_F \vb{e_y}$ and $-k_F \vb{e_y}$. To relax these electrons directly into an isotropic distribution (orientational relaxation), the momentum $q \approx 2k_F$ has to be exchanged by a phonon. In typical noble metals the momentum $2k_F > q_D$ exceeds the Debye sphere of a phonon and corresponds therefore to an umklapp process. In contrast, for normal processes a momentum $q < q_D$ is exchanged, so that multiple scattering events are required to achieve the momentum exchange of $2k_F$ and the relaxation proceeds accordingly slower. However, we note that the quantitative ratio of the contribution of normal and umklapp processes is also affected by the momentum-dependence of the electron-phonon matrix element in Eq.~\eqref{eq:def_eph_coupling}.
\\
In linear optics, the optical dielectric function of noble metals in the infrared and optical range has been the subject of several studies using ellipsometric measurements of thin films \cite{johnson_optical_1972, olmon_optical_2012, ordal_optical_1987, bennett_colloquium_1965, weaver_optical_1981,theye_investigation_1970, blanchard_highresolution_2003} and is therefore suitable for a comparison of the orientational relaxation time. In these studies, Drude relaxation times (as shown above they correspond to the momentum orientational relaxation time $\tau_{or}$) between $9$~fs and $25$~fs are measured: This broad range of different values results from specific sample properties such as surface roughness and impurities. A comparison with these results, delimited by the shaded grey areas in Fig.~\ref{fig:susceptibility} with data from Ref.~\cite{weaver_optical_1981, johnson_optical_1972}, prove that including umklapp processes to the electron-phonon scattering dynamics in necessary to obtain realistic results for the orientational relaxation time in noble metals. Again, for better illustration, we used a Drude fit instead of the original experimental data. 
For our simulation, we applied an effective mass $m$ in the electron dispersion derived from the lowest depicted data set \cite {johnson_optical_1972} in $\Re{\chi}$ which is reproduced in very good agreement by our numerical data.
\\
Our numerical results predict an orientational relaxation time in the lower limit of the experimental data explained by the neglected defect scattering and electron-electron scattering. However, already a qualitative and close to quantitative picture can be obtained by electron-phonon interaction: For electron-electron interaction, momentum can be exchanged within the electronic system, but the total momentum has to be conserved except for a reciprocal lattice vector. Thus, momentum relaxation is then only possible via umklapp processes. Furthermore, the electron-phonon interaction dominates the momentum orientational relaxation compared to the electron-electron umklapp scattering. The reason for this is that the squared matrix element of the electron-phonon interaction is proportional to $q^{-2}$ (cf.~Eq.~\eqref{eq:def_eph_coupling}) whereas the matrix element of the electron-electron interaction is proportional to $q^{-4}$, leading to a weaker interaction strength at large momentum exchange $q$ which holds for umklapp processes \cite{czycholl_solid_2023, ziman_electrons_1960, lawrence_electron-electron_1973}. 
\\
\textit{Ab initio} calculations on the orientational relaxation time by electron-phonon scattering also including umklapp processes \cite{brown_nonradiative_2016, mustafa_ab_2016} yield $24.0$~fs to $26.3$~fs in agreement with our results.
\\
In the linear regime, when the absorbed energy is negligible compared to the thermal broadening, a discussion of thermalization and cooling is not required.
\subsubsection{\label{sec:IR_nonlinear} Non-linear Excitation Regime}
In the non-linear regime, i.e.~at excitation with strength $E_0$ (see introduction of this Sec.~\ref{sec:IR}), besides orientational relaxation, also thermalization and cooling occur as they constitute energy relaxation mechanisms. In the following, the dependence of these three relaxation processes on the applied field strength or deposited energy, respectively, is discussed.
\\
We start with a brief comment on the transmitted field: As derived in Eq.~\eqref{eq:derivation_gamma_rad}, the strength of the self-interaction of the total radiation field via the current density depends only on the total electron density, which is conserved in a spatially homogeneous system and therefore cannot contribute to a non-linear response. Thus, the only source for non-linearities in our description are the Pauli blocking terms in the electron-phonon scattering in Eq.~\eqref{eq:scattering_eq}. 

\begin{figure}
    \centering
    \includegraphics{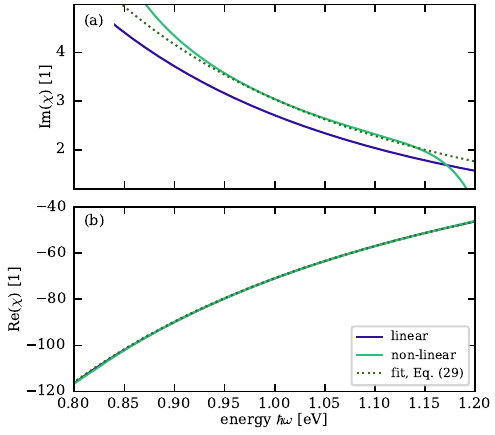}
    \caption{The (a) imaginary part of the optical susceptibility $\chi$ reveals an orientational relaxation time depending on the strength of excitation. In contrast, the (b) real part is mostly unaffected by the field strength.}
    \label{fig:nl_susceptibility}
\end{figure}

The non-linear susceptibility spectrum for excitations in the infrared regime is calculated in Fig.~\ref{fig:nl_susceptibility} (solid green lines) from the numerical results in the same way as in the linear regime in Eq.~\eqref{eq:susceptibility}:
\begin{align}
    \hat{\vb{j}}_s(\omega) = -i\omega \varepsilon_0 \chi^{nl}(\omega) \hat{\vb{E}}^{tot}(\omega) ~. \label{eq:nl_susceptibility}
\end{align}
Comparing this with the result from the linear regime (blue lines) reveals no changes for the real part but an increase of the imaginary part of the susceptibility. This suggests, in connection with Eq.~\eqref{eq:drude_limits} and our observations in Ref.~\cite{grumm_ultrafast_2025}, to assume an excitation dependent orientational relaxation in the non-linear regime.
\\
To discuss the orientational relaxation after non-linear excitation, we separate the orientational relaxation rate into linear and field-dependent non-linear contributions
\begin{align}
    \tau^{-1}_{or} = \tau^{-1}_{or, lin} + \tau^{-1}_{or, nl}(E^{tot}(t)) ~. \label{eq:def_nl_ort}
\end{align}
By expanding the non-linear contribution in powers of the total radiation field
\begin{align}
    \tau^{-1}_{or, nl}(E^{tot}(t)) = \sum_{n=1}^\infty (\tau^{-1}_{or,nl})^{(n)} (\Tilde{E}^{tot}(t) e^{i\omega_L t} +c.c.)^n ~, \label{eq:nl_ort}
\end{align}
one can connect the orientational relaxation rate to the total radiation field via a high order dissipative Kerr-type non-linearity $\tau_{or,nl}^{-1}(|\Tilde{E}^{tot}|^2)$ at the frequency of the incident field $\omega=\omega_L$. This means, the non-linear optical susceptibility $\chi \rightarrow \chi^{nl}(|\Tilde{E}^{tot}|^2)$ (and therefore also the complex refractive index of the film) becomes modulated as function of the intensity $|\Tilde{E}^{tot}|^2$ of the total radiation field. In a narrow frequency interval around the carrier frequency of the external field $\omega \approx \omega_L$, a Drude-like susceptibility results
\begin{align}
    \chi_{D}^{nl}&(\omega)|_{\omega \approx \omega_L} = \Tilde{\varepsilon}_b - \frac{\omega_p^2}{\omega^2+i\omega \left[ \tau^{-1}_{or, lin} + \tau^{-1}_{or, nl}(|\Tilde{E}^{tot}|^2) \right]} ~. \label{eq:nl_drude}
\end{align}
Thus, to determine the high order dissipative Kerr-type non-linear correction $\tau_{or, nl}(|\Tilde{E}^{tot}|^2)$ to the orientational relaxation time the same method as in Sec.~\ref{sec:IR_linear} can be applied but has to be limited to a narrow interval around $\omega_L$ (here, $\hbar \omega_L\pm 0.01$~eV is applied). This leads to $\tau_{or, nl}(|E_0|^2) = 155.0$~fs or $\tau_{or}(|E_0|^2) = 17.4$~fs, cf.~Eq.~\eqref{eq:def_nl_ort}, respectively, where the linear orientational relaxation time $\tau_{or, lin}$ is identical to Sec.~\ref{sec:IR_linear}. Fig.~\ref{fig:nl_susceptibility} depicts the fitted Drude-like susceptibility (dotted line) from Eq.~\eqref{eq:nl_drude}. As expected, the Drude-like susceptibility can be applied to derive the non-linear orientational relaxation time from the numerical result, but does not provide a model for the full frequency range. Since the real part is, due to its dominant singularity, Eq.~\eqref{eq:drude_limits}, mostly independent of the orientational relaxation time, no non-linear effects are observed there.
\\
Further numerical investigations confirm that the non-linear orientational relaxation time depends mainly on the amplitude of the external field $E_0$ but less on the temporal pulse duration. For excitations with $\hbar \omega_L=1$~eV, we observe that the orientational relaxation time is mostly independent from the pulse duration for pulses longer than the temporal oscillation period of the field ($FWHM \gg 2\pi/\omega_L$).
\\
Consequently, for an accurate description of non-linear effects in metallic structures, modifications of the imaginary part of the optical response due to microscopic many-body effects, here dominantly Pauli blocking, of $10$ to $20$\% can be expected. Non-linear hydrodynamic models \cite{cox_analytical_2017, scalora_second-_2010, ciraci_origin_2012}, such as those used for the description of plasmonic nanostructures, typically neglecting Pauli blocking, must therefore also be evaluated under the aspect of an intrinsic excitation dependent material absorption within changes of about $10~\%$ of the linear absorption. However, the impact of this non-linearity varies depending on the specific optical observable. While the transmission in the infrared investigated here is not very sensitive to the orientational relaxation time, Eq.~\eqref{eq:T,R_IR} below, other observables, e.g.~in a pump-probe scheme, may depend more on the non-linearity:
\\
To illustrate this, we recapitulate the pulse propagation studies from Fig.~\ref{fig:IR_spectrum}a,b. There, dotted lines depict the non-linear excitation in the infrared regime and indicate that the pulse propagation itself is basically unaffected by microscopic non-linearities in the electron-phonon interaction. This observation is based on the fact that the non-linear optical response in the infrared is expressed by a field-dependent orientational relaxation time (cf.~Eq.~\eqref{eq:nl_ort}), which leaves the transmission and reflection coefficients, Eq.~(\ref{eq:T}, \ref{eq:R}), unaffected in the limit $\omega^2 \gg \tau_{or}^{-2}$
\begin{align}
    T \approx \Big(\frac{2c_0 n_{out}}{d \omega_p^2} \Big)^2 \omega^2 ~~,~ R \approx 1 - \Big( \frac{2 c_0 n_{out}}{d\omega_p} \Big)^2 \omega^2 ~. \label{eq:T,R_IR}
\end{align}
As we will see in Sec.~\ref{sec:THz}, this is not the case for THz excitations where the impact of the orientational relaxation time on the optical response is much stronger (cf.~Eq.~\eqref{eq:drude_limits}). 

\begin{figure*}
    \centering
    \includegraphics{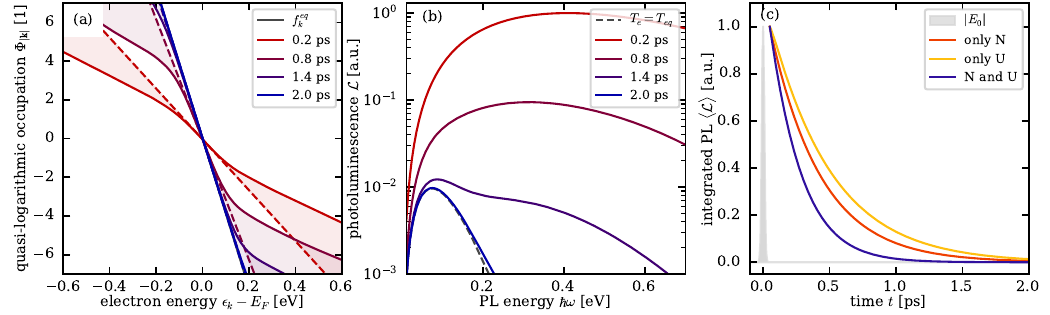}
    \caption{In (a), the computed energy-resolved non-equilibrium electron distribution (solid lines) can be separated in thermal (dashed lines) and non-thermal distributions (shaded areas). The corresponding photoluminescence spectra in (b) depend on the thermal and non-thermal energy of the electron gas and decrease in intensity while the electron gas dissipates energy to the phonon bath. In thermodynamic equilibrium (dashed line) the spectrum corresponds to Planck's radiation law. The temporal decay of the spectrally integrated photoluminescence in (c) is related to an energy relaxation time $\tau_{er} = 0.24~$ps (blue line). (c) also compares the dynamics analyzed in terms of normal (N) and umklapp (U) processes.} 
    \label{fig:luminesence}
\end{figure*}

In the next step, we show that based on the full momentum-resolved Boltzmann equation \eqref{eq:thin_film_boltzmann} known effects of energy relaxation can be recovered and we connect them to experimental observables. In contrast to the orientational relaxation, which is completed after about $0.1$~ps, the dissipation of energy to the phonon bath proceeds on a significantly longer time scale up to several picoseconds. The process of energy relaxation is divided into thermalization and cooling being only relevant after non-linear excitation when a significant amount of energy is absorbed by the electron gas ($\epsilon_{abs} \gtrsim k_B T_{eq}$). Thermalization describes the transition of a non-thermal electron distribution into a thermal Fermi-Dirac distribution. This process is mainly driven by electron-electron scattering \cite{del_fatti_nonequilibrium_2000, seibel_time-resolved_2023} but also supported by electron-phonon scattering \cite{rethfeld_ultrafast_2002, mueller_relaxation_2013, grumm_ultrafast_2025}. In contrast, energy transfer from the electronic to the phononic system results in cooling of the electron gas, where the energy of hot thermal electrons is dissipated into the phonon bath until a thermodynamic equilibrium is reached \cite{grumm_ultrafast_2025}. Studying both processes exclusively driven by electron-phonon scattering, i.e.~the limit of not too strong excitations, as done here, they proceed on a similar time scale and are hard to disentangle \cite{mueller_relaxation_2013}. 
\\
Since the radiative self-interaction (cf.~Eq.~(\ref{eq:total_radiation_field}, \ref{eq:thin_film_boltzmann})) is directly related to the current density, its influence is only important during the orientational relaxation time, when finite currents $\vb{j}_s$ exist. Thus, re-radiation after ultrashort optical excitation only influence the built-up of the initial conditions for thermalization and cooling, but not their energy relaxation dynamics.
\\
The computed energy-resolved non-equilibrium electron distribution for selected times after the excitation with strength $E_0$ at $t=0$ and the followed energy relaxation is depicted in Fig.~\ref{fig:luminesence}a (solid lines) in a quasi-logarithmic representation $\Phi_{|\vb{k}|}$. As discussed in Ref.~\cite{grumm_ultrafast_2025, mueller_relaxation_2013}, the dashed line represents the thermal contribution to the electron distribution, 
whose slope is linked to the temperature of the thermal electrons. However, this temperature is not a function of state of the entire electron system as long as non-thermal electrons exist \cite{puglisi_temperature_2017, dubi_hot_2019}. This non-thermally distributed electrons are visualized by the shaded areas.
Thus, thermalization is connected to a decreasing deviation between the full and the thermal electron distribution while cooling is described by the increasing slope of the thermal distribution in time. Both processes, driven by electron-phonon interaction, proceed on the same time scale about more than $1$~ps \cite{rethfeld_ultrafast_2002, mueller_relaxation_2013}.
\\
To connect our theoretical calculations to observables, we study photoluminescence spectra: Photoluminescence occurs via the photon emission in the course of radiative recombination of excited hot (thermal and non-thermal) electrons and is therefore suitable to track the energy dissipation dynamics. When considering the photoluminescence of metals, a distinction between inter- and intraband processes is necessary \cite{boyd_photoinduced_1986, shahbazyan_theory_2013, bowman_quantum-mechanical_2024, rodriguez_echarri_nonlinear_2023}. In this work, we focus on intraband processes comparable to Ref.~\cite{ono_ultrafast_2020, sivan_theory_2021, loirette-pelous_theory_2024}. 
\\
The photoluminescence $\mathcal{L}(\omega,t)$ is defined via the product of the density of photon and electron states participating at recombination \cite{ono_ultrafast_2020, sivan_theory_2021}
\begin{align}
    \mathcal{L}(\omega,t) \propto& \omega^2 \int_{-\infty}^\infty \mathrm{d}\epsilon_\mathbf{k} \int_{-\infty}^\infty \mathrm{d}\epsilon_\mathbf{k'} \times \label{eq:def_pl}  \\
    & \times \sqrt{\epsilon_\mathbf{k}\epsilon_\mathbf{k'}} f_\mathbf{k}(t) (1-f_{\mathbf{k'}}(t)) \delta(\epsilon_\mathbf{k'}-\epsilon_\mathbf{k}+\hbar \omega)  ~. \nonumber
\end{align}
Eq.~\eqref{eq:def_pl} includes the occupation $f_\mathbf{k}$ and Pauli blocking contribution $(1-f_{\mathbf{k'}})$ of the radiatively recombining states $\vb{k}$ and $\vb{k'}$ (or electron and hole contribution) as well as the density of states in the integrand. 
A comparison with the internal energy density $u_e = \frac{1}{\pi^2} \int_{-\infty}^\infty \mathrm{d}\epsilon_\mathbf{k} \epsilon_\mathbf{k} f_{\vb{k}}$ reveals, that the photoluminescence is closely related to the energy of the electron gas. In fact, Eq.~\eqref{eq:def_pl} is the convolution of the energy density of occupied states with unoccupied states. In the case of purely thermal electrons, i.e.~$f_\mathbf{k}$ is given by a Fermi-Dirac distribution, and in the limit $\hbar \omega \ll E_F$, the photoluminescence formula \eqref{eq:def_pl} results in Planck's radiation law for the emitted photon number \cite{ono_ultrafast_2020, haug_hot-electron_2015}. 
\\
The photoluminescence spectra, calculated with Eq.~\eqref{eq:def_pl} from the distributions in Fig.~\ref{fig:luminesence}a, are shown in Fig.~\ref{fig:luminesence}b, where the dashed line is the photoluminescence at the equilibrium temperature and is equal to a spectrum described by Planck's radiation law. The influence of non-thermal electrons is still visible $1.4$~ps after the excitation (purple line): While the photoluminescence emission maximum agrees with the thermodynamic equilibrium (dotted line), the tail in the spectrum ($\hbar \omega > 0.2$~eV) can be related to the photoluminescence of non-thermal electrons. The following electron-phonon thermalization persists until more than $2$~ps after excitation.
\\
To access the energy relaxation dynamics from the photoluminescence dynamics as function of time, Fig.~\ref{fig:luminesence}c studies the spectrally integrated photoluminescence $\langle \mathcal{L} \rangle (t) = \int\mathrm{d}\omega ~\mathcal{L}(\omega,t)$ derived from the spectra in Fig.~\ref{fig:luminesence}b. The resulting decay incorporates both processes of cooling and thermalization and is connected to an energy relaxation time of $\tau_{er} = 0.24$~ps.
Fig.~\ref{fig:luminesence}c also analyzes the cooling dynamics in terms of normal and umklapp processes: Considering only normal processes reveals an energy relaxation time of about $\tau_{er}^N = 0.38$~ps. Here, the contribution of umklapp processes $\tau_{er}^U = 0.47$~ps is one fourth slower than the normal process contribution. Thus, in contrast to orientational relaxation, where umklapp relaxation is three times faster than the corresponding normal process relaxation (cf.~Fig.~\ref{fig:susceptibility}), the impact of umklapp processes to energy relaxation is much weaker. This can be understood as follows: Since energy relaxation is connected to the energy exchange between electrons and phonons and the possible energy carried by phonons of several meV does not differ between normal and umklapp processes, none of the two processes is preferred. However, since the electron-phonon coupling is weaker for umklapp processes (cf.~Eq.~\ref{eq:def_eph_coupling}), normal processes dominate energy relaxation.
\\
In agreement with other studies \cite{fann_direct_1992, brown_experimental_2017, giri_experimental_2015, suemoto_relaxation_2021}, we observe a non-linearly increasing energy relaxation time for higher external field strengths. However, experimental studies based on the luminescence or time domain transient reflection measurements as well as theory models detect energy relaxation times between $0.8$~ps and $1.2$~ps in gold \cite{sun_femtosecond_1993, sun_femtosecond-tunable_1994, groeneveld_femtosecond_1995, del_fatti_nonequilibrium_2000, giri_experimental_2015, suemoto_observation_2019, suemoto_relaxation_2021}. 
In comparison with other theoretical studies on gold \cite{del_fatti_nonequilibrium_2000, rethfeld_ultrafast_2002, mueller_relaxation_2013, brown_experimental_2017}, our description shows a faster energy relaxation. In our description, the total radiation field $E^{tot}$, acting as driving force on the electrons, is reduced in amplitude against the external field by approximately a factor of four (cf.~Fig.~\ref{fig:IR_spectrum}) due to the self-consistent treatment. This results in less strong pronounced non-equilibrium electron distributions in the thin film. 
This is in agreement with our observation provided in Ref.~\cite{grumm_ultrafast_2025} for a bulk system without radiative self-interaction: There, we obtained an energy relaxation dominated by cooling in agreement with the above cited theory and experimental studies for an external field even three times weaker than in the present work.
\subsection{\label{sec:THz} THz Excitations}
\begin{figure}
    \centering
    \includegraphics{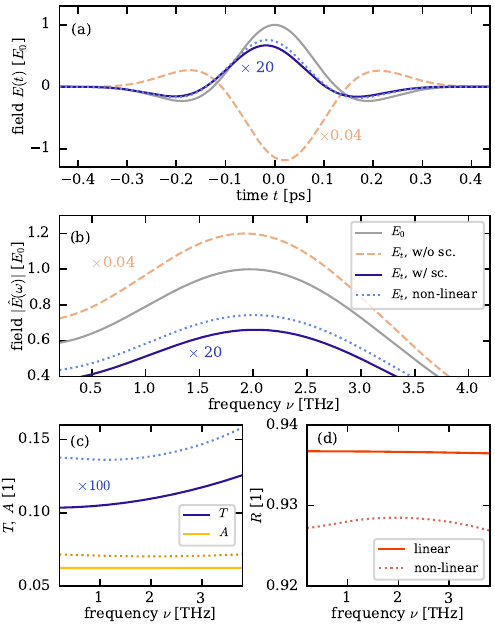}
    \caption{The incident THz field $E_0$ and transmitted field $E_t$ calculated with and without self-consistency (sc., solid or dashed lines) in the thin film Boltzmann equation \eqref{eq:thin_film_boltzmann} are shown in (a) time- and (b) frequency-domain. A linear field strength $E_0=10^{-3}~\mvcm$ is applied. In the non-linear regime, shown here exemplarily for $E_0=4~\mvcm$, the transmitted field (dotted line) increases non-linearly in amplitude. In (c,d) the calculated transmission, reflection and absorption ($T,~R,~A$) are shown for linear (solid lines) and non-linear excitations (dotted lines).}
    \label{fig:THz_spectrum}
\end{figure}

In the previous section \ref{sec:IR}, the optical response of a thin gold film after infrared excitation was discussed, where the orientational relaxation rate $\tau_{or}^{-2} \ll \omega_L^2$ is much smaller than the excitation frequency. Therefore, the imaginary part of the susceptibility, Eq.~\eqref{eq:drude_limits}, is dominantly sensitive for changes in the orientational relaxation but not the transmitted field or the transmission itself (cf.~Eq.~\eqref{eq:T,R_IR}).
\\
In contrast, in this section, the focus is on the case where the excitation frequency is at least on the same time scale as the relaxation dynamics and strong momentum-polarized non-equilibrium electron distributions are built-up during the competing relaxation processes.
This corresponds to excitations in the THz regime, where the temporal oscillation period of the incident field is much longer than the orientational relaxation time ($\omega_L^2 \ll \tau_{or}^{-2}$). In this regime, the induced momentum-polarized non-equilibrium electron gas exists over a period of hundreds of fs up to ps. As we will show in this section, this activates a THz field-induced Pauli blocking non-linearity of the electron-phonon interaction which is directly observable in the transmission signal.
We predict, that the non-linearity manifests as a high order dissipative Kerr-type non-linear self-modulation of the propagating THz field for field strengths above $0.1~\mvcm$ at $\nu_L = 2$~THz ($\hbar \omega_L=8$~meV). Here, we apply external THz field strengths ranging from $E_0=10^{-3}~\mvcm$ in the linear regime to $E_0=4~\mvcm$ in the non-linear regime. Again, as in Sec.~\ref{sec:IR}, both regimes are separated by the relation of the absorbed energy $\epsilon_{abs}$ and the thermal broadening of the equilibrium electron distribution $k_B T_{eq}$. Such high-intense THz fields became possible by versatile techniques in recent years \cite{gollner_highly_2021, koulouklidis_observation_2020, nicoletti_nonlinear_2016, dhillon_2017_2017, leitenstorfer_2023_2023}. Today, a frequency range between $0.1$~THz and $30$~THz and field strengths up to $100~\mvcm$ \cite{koulouklidis_observation_2020} in single-shot pulses and at least $10~\mvcm$ for repeatable, comparable stable single-cycle pulses \cite{gollner_highly_2021} can be generated.
\\
In the following, we discuss numerical results for a $d=5$~nm thin gold film under excitation with a $\nu_L = 2$~THz pulse with a temporal pulse duration of $FWHM = 300$~fs. This pulse duration includes approximately one and a half period of the THz oscillation (cf.~Fig.~\ref{fig:THz_spectrum}) similar to typical pulses generated in experiments \cite{koulouklidis_observation_2020, gollner_highly_2021} and allows a strong momentum-polarization of the electron gas. Thin films of several nm thickness could be produced by sputtering or in a chemically controlled process similar to other metal nanostructures \cite{staechelin_size-dependent_2021} and are still transmitting at THz frequencies.
\\
This section focus on the discussion of the pulse propagation and orientational relaxation for non-linear THz excitations but will not provide a discussion of thermalization and cooling as it occurs in a similar fashion as in the infrared regime in Sec.~\ref{sec:IR_nonlinear}. 
\subsubsection{\label{sec:THz_linear} Linear Excitation Regime}
Fig.~\ref{fig:THz_spectrum} depicts the calculated transmitted THz field in (a) time- and (b) frequency-domain for linear excitations (solid lines). Here, as in the infrared regime in Fig.~\ref{fig:IR_spectrum}, the transmitted field is phase shifted and reduced in amplitude against the incident field (grey). However, the amplitude reduction is more than ten times stronger than in the infrared regime resulting in a ten times weaker total radiation field $\vb{E}_t = \vb{E}^{tot}$ in Eq.~\eqref{eq:total_radiation_field} acting on the conduction electrons than for infrared pulses with comparable field strength. 
This feature is connected with the dominating imaginary part of the optical susceptibility, Eq.~\eqref{eq:drude_limits}, which leads to the well-known strong reflection and low transmission of noble metals, as shown in Fig.~\ref{fig:THz_spectrum}c,d (solid lines). We note, that neglecting the self-consistency in the theory in the THz regime would lead to non-physical results for the transmitted field (dashed lines in Fig.~\ref{fig:THz_spectrum}a,b) with an amplitude larger than the external field. 
\\
The orientational relaxation time in the linear regime is again determined via fitting the calculated optical response with a Drude model. However, the linear approximation used to derive Eq.~\eqref{eq:im_drude_chi} is not valid for THz frequencies. Here, the linear transmission spectrum depicted in Fig.~\ref{fig:THz_spectrum}c is connected via Eq.~\eqref{eq:T} with the full Drude susceptibility in Eq.~\eqref{eq:chi_drude}. This fitting procedure results in an orientational relaxation time of $\tau_{or}=19.5~$fs almost identical with the value derived in the infrared regime (cf.~Sec.~\ref{sec:IR_linear}).
\subsubsection{\label{sec:THz_nonlinear} Non-Linear Excitation Regime}
\begin{figure}
    \centering
    \includegraphics{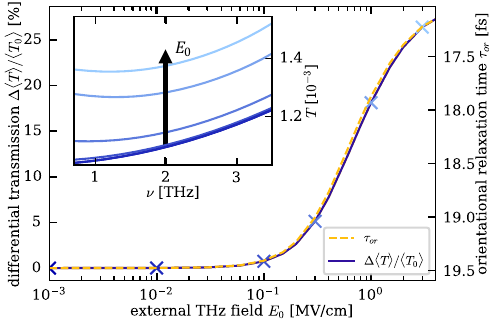}
    \caption{The inset shows the transmission spectra $T$ for selected external THz field strengths (markers in the main plot). The increasing differential transmission in the main plot (solid line) reveals the many-body non-linearity in the THz regime. The corresponding orientational relaxation time on the right axis (note the reverse axis) agrees with the non-linear trend of the differential transmission.}
    \label{fig:THz_ort_vs_T}
\end{figure}

\begin{figure}
    \centering
    \includegraphics{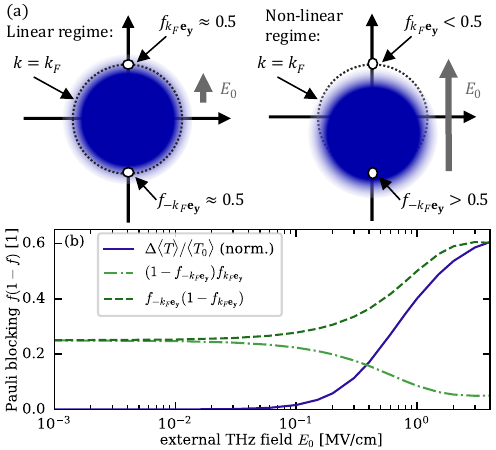}
    \caption{In (a) the electron distribution (Fermi sphere, blue) with finite temperature $T_e$ is sketched in presence of an external field $E_0$ in the linear (left) and non-linear regime (right). The dotted line represents the Fermi level in equilibrium. In (b) the Pauli blocking terms $f(1-f)$ (evaluated at $\pm k_F \vb{e_y}$, small circles in (a)) are analyzed as function of the external field strength. For comparison, the normalized differential transmission from Fig.~\ref{fig:THz_ort_vs_T} is shown (solid line).}
    \label{fig:pauli_blocking}
\end{figure}

Studying in Fig.~\ref{fig:THz_spectrum}a,b the transmitted field for non-linear excitations (dotted lines) reveals a relative increase in amplitude and an additional phase modulation compared to the linear regime. This phase modulation causes a weak red-shift of the transmitted field in the non-linear regime, which is less than $1$\% of the THz frequency $\nu_L$. In contrast, the amplitude of the transmitted non-linear field increases with more than $10$\% significantly. This non-linearity appears in Fig.~\ref{fig:THz_spectrum}c,d by an increase of transmission and absorption and a decrease of the thin film reflection. We will show in the following, that this is caused by the non-linear field strength dependent orientational relaxation time already observed for infrared excitations (cf.~Sec.~\ref{sec:IR_nonlinear}). However, in contrast to the infrared regime, the impact of the orientational relaxation on the transmission is more pronounced in the THz regime. Qualitatively, this behavior can be understood by applying a non-linear Drude-like susceptibility as in Eq.~\eqref{eq:nl_drude} to the transmission and reflection in Eq.~(\ref{eq:T}, \ref{eq:R}). The corresponding transmission and reflection in the limit of small frequencies ($\omega_L^2 \ll \tau_{or}^{-2}$) reads
\begin{align}
    &T \approx \Big(\frac{2c_0 n_{out}}{d \omega_p^2} \Big)^2 \tau_{or}^{-2}(|E_0|^2) ~~, \nonumber \\
    &R \approx 1 - \frac{4 c_0 n_{out}}{d\omega_p} \tau_{or}^{-1}(|E_0|^2) ~, \label{eq:T,R_THz}
\end{align}
in agreement with the observations in Fig.~\ref{fig:THz_spectrum}c,d assuming a non-linearly increasing orientational relaxation time $\tau_{or}$.
Thus, the relative impact of a non-linear orientational relaxation is strongest for the transmission. Therefore, we choose this as an observable to discuss the non-linearity and its connection with orientational relaxation in the following.
To quantify the non-linear increase of transmission, we define the differential transmission 
\begin{align}
    \Delta \langle T \rangle/\langle T_0 \rangle = \frac{\langle T \rangle - \langle T_0 \rangle}{\langle T_0 \rangle} \label{eq:diff_transmission}
\end{align}
via the linear transmission $\langle T_0 \rangle$. Here, the spectrally integrated transmission is introduced
\begin{align}
    \langle T \rangle = \frac{\int \mathrm{d}\omega ~ \vert \hat{E}_t(\hat{E}^{tot}, \omega)\vert^2}{\int \mathrm{d}\omega ~ \vert \hat{E}_0(\omega)\vert^2} ~,
\end{align}
where in the non-linear regime $\langle T \rangle$ depends on the total radiation field $\hat{E}^{tot}$ but becomes field independent in the linear regime $\langle T \rangle = \langle T_0 \rangle$. Obviously, the differential transmission, Eq.~\eqref{eq:diff_transmission}, is zero in the linear regime and deviations from zero indicate the non-linear regime. Fig.~\ref{fig:THz_ort_vs_T} displays the differential transmission as function of the external THz field strength. For external field strengths below $E_0=0.1~\mvcm$ the transmission behaves linear. This value can be considered as rough boundary between the linear and non-linear regime. Physically, this value is connected to a total radiation field $E^{tot}(t)$ which corresponds to an absorbed energy of $\epsilon_{abs} = 90$~meV in the order of the thermal broadening $k_B T_{eq}$. Above this threshold, i.e.~for higher field strengths, the transparency of the thin film starts to increase and shows a saturation behavior. Above $E_0=2~\mvcm$ the differential transmission reaches a maximum at about $28$\%. 
\\
The non-linearity is connected with the field-strength dependent decrease of the orientational relaxation time: As previously mentioned in Sec.~\ref{sec:IR_nonlinear} to determine the high order dissipative Kerr-type non-linear orientational relaxation time, we assume a Drude-like susceptibility, Eq.~\eqref{eq:nl_drude}, in a narrow frequency interval around the THz frequency $\nu_L$ (here $\nu_L \pm 0.02~$THz is applied). Thus, the orientational relaxation time, depicted in Fig.~\ref{fig:THz_ort_vs_T} (dashed line), is obtained by fitting the numerically calculated transmission data, inset in Fig.~\ref{fig:THz_ort_vs_T}, with Eq.~(\ref{eq:T}, \ref{eq:nl_drude}). In the linear regime, the orientational relaxation time is $\tau_{or}=19.5~$fs, decreases for non-linear excitation above $0.1~\mvcm$, and saturates at $\tau_{or}(|E_0|^2)=17.1~$fs for non-linear field strengths (here $E_0=4~\mvcm$). This agrees with the trend of the differential transmission including the saturation at high field strengths. 
\\
Microscopically, this behavior can be understood by analyzing the momentum-dependent Pauli blocking terms $f_\mathbf{k} (1-f_\mathbf{k+q+G})$ (and the counteracting term with $\vb{k} \leftrightarrow \vb{k+q+G}$) in the electron-phonon scattering equation \eqref{eq:scattering_eq}. We provide a qualitative picture which considers only one momentum state and does not include the competing relaxation dynamics proceeding simultaneously with excitation:  
Consider a scattering process, as illustrated in Fig.~\ref{fig:pauli_blocking}a, in which the initial and final states of the scattering electron are on opposite sides of the Fermi sphere parallel to the polarization direction of the external field, i.e.~$\vb{k} = +k_F \vb{e_y}$ and $\vb{k+q+G} = -k_F \vb{e_y}$. Then, in thermodynamic equilibrium or for linear excitations, where the electron distribution is only weakly perturbed, is $f_{\pm k_F \vb{e_y}}\approx 0.5$ and the Pauli blocking terms become $(1-f_{-k_F \vb{e_y}}) f_{k_F \vb{e_y}} = f_{-k_F \vb{e_y}} (1-f_{k_F \vb{e_y}}) = 0.25$. With increasing field strength above of the onset of the non-linearity at $E_0=0.1~\mvcm$, the Fermi sphere is shifted along the polarization direction (cf.~Fig.~\ref{fig:pauli_blocking}a right) and the value of the Pauli blocking term changes. In total, this leads to an increase of the electron-phonon scattering rate (see discussion of Fig.~\ref{fig:pauli_blocking}b below) resulting in an increase of the orientational relaxation rate and transmission (cf.~Eq.~\eqref{eq:T,R_THz}). 
However, fermions have to obey $0 \leq f_\mathbf{k} \leq 1$. This results in $(1-f_{-k_F \vb{e_y}}) f_{k_F \vb{e_y}} \geq 0$ or $f_{-k_F \vb{e_y}}(1-f_{k_F \vb{e_y}}) \leq 1$, respectively, explaining the saturation of the orientational relaxation time at high field strengths. 
This behavior is verified by our numerical results in Fig.~\ref{fig:pauli_blocking}b: Here, both Pauli blocking terms are calculated as function of the external field strength and are evaluated during the THz excitation pulse at the time with maximum momentum-polarization at $t \approx -0.1$~ps (cf.~maximum of $\Delta f_\mathbf{k_y}$ in Fig.~\ref{fig:distribution}b). While in the linear regime both Pauli blocking terms appearing in the scattering equation \eqref{eq:scattering_eq} are identical, they split for non-linear excitations. However, due to the competing relaxation dynamics, they do not reach the maximum saturation values at $0$ or $1$ for very high field strengths, but already saturate at about $0.05$ or $0.60$, respectively. This leads in total to the fact that the sum of both Pauli blocking terms increases from $(1-f_{-k_F \vb{e_y}}) f_{k_F \vb{e_y}} + f_{-k_F \vb{e_y}} (1-f_{k_F \vb{e_y}}) = 0.5$ in the linear regime to a value $> 0.5$ for non-linear excitations resulting in an increased electron-phonon scattering rate.
This discussion can be applied analogously to the non-linear field strength dependence of the orientational relaxation time in the infrared regime in Sec.~\ref{sec:IR_nonlinear} and does not depend on the time scale of excitation.

\begin{figure*}
    \centering
    \includegraphics{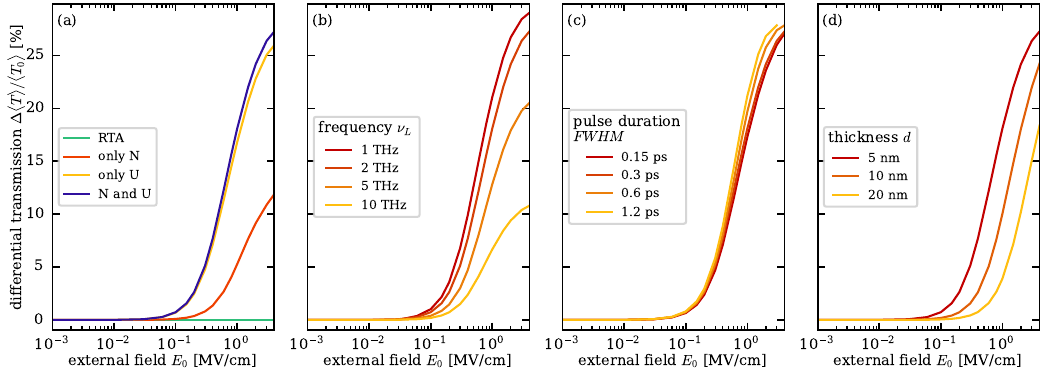}
    \caption{In (a) the non-linearity is compared for normal (N) and umklapp (U) processes in the electron phonon interaction as well as for a relaxation time approximation (RTA). In (b), (c) and (d) the THz frequency, pulse duration or film thickness are varied, respectively. All other parameters in each figure are set as described in the introduction of this section.}
    \label{fig:THz_NL}
\end{figure*}

To gain further insights in the non-linearity as function of the external THz field strength $E_0$ and its connection with the orientational relaxation, Fig.~\ref{fig:THz_NL} depicts in (a) the dependence on the complexity of electron-phonon scattering theory and discusses the non-linearity for varying (b) THz frequencies, (c) pulse durations, and (d) film thicknesses.
\\
Fig.~\ref{fig:THz_NL}a compares different approaches for electron-phonon relaxation. In the relaxation time approximation, as assumed in the phenomenological Drude model (cf.~Eq.~\eqref{eq:rta_boltzmann}), the Pauli blocking terms in the scattering equation \eqref{eq:scattering_eq} are linearized and expressed via constant equilibrium Fermi-Dirac distributions $f_\mathbf{k}(t) (1-f_\mathbf{k+q+G}(t)) \rightarrow f_\mathbf{k}(t) (1-f_\mathbf{k+q+G}^{eq})$ (and vice versa). Thus, the orientational relaxation can not be affected by changes in the Pauli blocking due to the strong momentum-polarization of the electron gas and no non-linearity emerges as verified by our results. Solving the full scattering equation \eqref{eq:scattering_eq} for only normal processes in the electron-phonon interaction reveals a weaker non-linearity in the differential transmission saturating already at $13$\%. In contrast, umklapp processes contribute with a saturation at $27$\% the main part to the non-linearity. This demonstrates that orientational relaxation is indeed the driving process of the non-linearity and not energy relaxation (cooling and thermalization), which depend less on umklapp processes (cf.~discussion in Sec.~\ref{sec:IR_nonlinear}).
\\
Fig.~\ref{fig:THz_NL}b supports this conclusion by comparing the non-linearity for various THz frequencies with a pulse duration of $FWHM = 300$~fs. The absorbed energy $\epsilon_{abs}$ for excitation with $\nu_L=10$~THz is approximately $1.5$ times larger than for $\nu_L=2$~THz. Nevertheless, the observed non-linearity is less pronounced for the case with higher absorbed energy. However, this can be explained by the fact that for longer temporal oscillation periods of the THz field (lower frequencies), the time on which the electron gas becomes momentum-polarized is longer during each oscillation half-cycle. Thus, a low frequency THz field induces
a stronger momentum-polarization in the electron gas and the Pauli blocking terms (and therefore the non-linearity) appear more pronounced than in the case with a higher frequency.
\\
Fig.~\ref{fig:THz_NL}c depicts the non-linearity in the differential transmission for selected pulse durations between $FWHM = 0.15$~ps and $1.2$~ps ($\nu_L = 2$~THz for all curves). Here, longer pulse durations are associated with larger absorbed energies, e.g.~the absorbed energy for the $1.2$~ps pulse is more than four times larger than for the $0.3$~ps pulse. However, the impact of the pulse duration on the non-linearity is only marginal. This is in agreement with the observations from Fig.~\ref{fig:THz_NL}b: The non-linearity relates less to the absorbed energy than to the excited momentum-polarization, which is independent of the pulse duration and depends entirely on the frequency and peak field strength. The small deviations between the different pulse durations for field strengths above $1.0~\mvcm$ in Fig.~\ref{fig:THz_NL}c can be addressed with the parallel proceeding thermalization and cooling relaxation, which, in contrast, depend on the absorbed energy \cite{fann_direct_1992, brown_experimental_2017, giri_experimental_2015, suemoto_relaxation_2021} and occurring on a longer ps time scale (cf.~Sec.~\ref{sec:IR_nonlinear}).
\\
Fig.~\ref{fig:THz_NL}d addresses the influence of the film thickness $d$ on the emerging non-linearity. As derived in Eq.~\eqref{eq:derivation_gamma_rad}, the radiative damping of the external field and thus the strength of the total radiation field, which excites the electrons into the non-equilibrium distribution, is approximately linearly related to the film thickness. Fig.~\ref{fig:THz_NL}d agrees with this: If we assume a film twice as thick, the onset of the non-linearity (measured in terms of the external field strength) and the general curve of the non-linearity shifts by a factor of two to higher field strengths. This is due to the fact that for the same total radiation field, which finally acts on the electrons, a approximately two-times stronger external field is required.
\\
In summary, Fig.~\ref{fig:THz_NL} demonstrates that umklapp processes of electron-phonon scattering contribute dominantly to the non-linear transmission in the THz range. In addition, the non-linearity depends less on the absorbed energy of the THz excitation than on the duration on which the momentum-polarization of the electron gas is maintained. Therefore, we attribute the high order dissipative Kerr-type non-linearity in the transmission of thin noble metal films in the THz range to a field strength dependent orientational relaxation rate generated by the Pauli blocking in the electron-phonon scattering terms in Eq.~\eqref{eq:scattering_eq}.
\section{\label{sec:conclusion} Conclusion}
\begin{table}
    \caption{\label{tab:summary}%
    Summary of the optical properties of thin noble metal films in the infrared and THz regimes.}\begin{ruledtabular}
        \begin{tabular}{lcc}
             & IR & THz \\
            \hline
            $\chi_D(\omega)$ &  $\propto - \frac{\omega_p^2}{\omega^2} + i \frac{\omega_p^2 \gamma_D}{\omega^3}$ & $\propto - \frac{\omega_p^2}{\gamma_D^2} + i \frac{\omega_p^2}{\omega\gamma_D}$ \\
            $T(\omega)$ & $\propto \omega^2$ & $\propto \tau_{or}^{-2}$ \\
            $2\pi/\omega_L$ & $\approx 4$ fs & $\approx 500$ fs \\
            $\tau_{or}$ (linear) & $19.6$ fs & $19.5$ fs \\
            $\tau_{or}(|E_0|^2)$ (non-linear) & $\gtrsim 17$ fs & $\gtrsim 17$ fs \\
            Pauli blocking $f(1-f)$ & yes & yes \\
            Is $\Delta f_\mathbf{k}(t) \propto E_0(t)$? & yes & no \\
            $\Rightarrow \Delta\langle T \rangle/\langle T_0 \rangle$ & $\approx 0$ \% & $>20$ \% \\
            \hline
            \textit{Non-linearity detectable} \\ \textit{in the transmission?} & no & yes \\
        \end{tabular}
    \end{ruledtabular}
\end{table}
We introduced a self-consistent framework of Maxwell's equations and the kinetic momentum-resolved Boltzmann equation to describe the dynamics of conduction electrons in thin noble metals films after infrared and THz excitations. By deriving an analytical solution of Maxwell's equations for a thin film, the numerical complexity of the problem is reduced, allowing for a momentum-resolved treatment of the electron dynamics. In this way, the characteristic relaxation processes of orientational relaxation, thermalization and cooling are addressed in terms of optical observables for thin films in the infrared and THz regime.
\\
By including both, normal and umklapp processes to electron-phonon interaction, an orientational relaxation time of $\tau_{or}= 19.6$~fs is determined in the linear regime in agreement with experimental and theoretical studies. We observe, that umklapp processes dominate the momentum orientational relaxation, whereas for energy relaxation  normal and umklapp processes contribute equally. 
\\
For non-linear excitations, a decrease of the orientational relaxation time of more than $10$\% is derived, which is caused by the Pauli blocking term in electron-phonon scattering activated for a strongly momentum-polarized non-equilibrium electron gas. A summary of the properties of the non-linearity in the infrared and THz regime as well as its effect on optical observables is provided in Tab.~\ref{tab:summary}.
Based on this, an optically detectable high order dissipative Kerr-type non-linearity in the THz regime is predicted: Since the temporal oscillation period of the THz field is longer than the typical orientational relaxation time, it is sensitive even for small changes in the orientational relaxation dynamics. Thus, for external THz field strengths above $0.1~\mvcm$, an increase of the thin film transmission by more than $20$\% is observed.
\begin{acknowledgments}
We acknowledge fruitful discussions with Robert Salzwedel, Lara Greten, Joris Strum, Leo Bretz, Robert Lemke (TU Berlin), Yannic Stächelin (Uni Hamburg), Rokas Jutas, Audrius Pugzlys (TU Wien) and Francesca Calegari (DESY Hamburg). \\
This work was supported by the German Science Foundation (DFG) under Grant 432266622.
\end{acknowledgments}
\appendix
\begin{table}
    \caption{\label{tab:parameters}
    Material parameters for simulation in gold.}\begin{ruledtabular}
        \begin{tabular}{lcr}
            Parameter & Value & Reference \\
            \hline
            $\varepsilon_\infty$ & $9.07$ & \cite{vial_improved_2005} \\
            $\varepsilon_{out}$ & $1.0$ &  \\
            $n_0$ & $59.0~\frac{1}{\text{nm}^3} $ & \cite{ashcroft_solid_1976} \\
            $m$ & $0.99~m_e$ & \cite{ashcroft_solid_1976} \\
            $a_0$ & $0.408$~nm & \cite{ashcroft_solid_1976} \\
            $E_F$ & $5.53 $~eV & \cite{ashcroft_solid_1976} \\
            $k_F $ & $ 12.0~\frac{1}{\text{nm}}$ & from $E_F= \frac{\hbar^2 k_F^2}{2m}$ \\
            $q_{D}$ & $15.2~\frac{1}{\text{nm}} $ & from $q_{D} = (6\pi^2 n_0)^{\frac{1}{3}}$ \\
            $\kappa$ & $16.9~\frac{1}{\text{nm}} $ & \cite{ashcroft_solid_1976}\\
            $c_{LA}$ & $3.24\cdot 10^ {-3}~\frac{\text{nm}}{\text{fs}} $ & \cite{lide_crc_2003} \\
        \end{tabular}
    \end{ruledtabular}
\end{table}
\section{\label{app:umklapp} Umklapp processes in fcc lattices}
Lets consider an electron-phonon interaction, as described in Eq.~\eqref{eq:hamiltonian}, where the electrons in the initial and final state have momenta $\vb{k}$ and $\vb{k}+\vb{q+G}$ in the first Brillouin zone. Then, the maximum momentum transfer $|\vb{q+G}|$ for a scattering process across the Fermi sphere is approximately $2k_F$ \cite{lawrence_umklapp_1972, ziman_electrons_1960}. In a reduced zone scheme, the phonon momentum is restricted to the Debye sphere $|\vb{q}| \leq q_D$. In particular one derives via the triangular inequality
\begin{align}
    |\vb{G}| \leq |\vb{q}| + 2k_F \leq q_D + 2k_F ~. \label{eq:max_G}
\end{align}
This estimate is even valid for a broadening of the Fermi edge up to $2.2$~eV in gold.
\\
For integer $l,m,n$ and a lattice constant $a_0$, the set of reciprocal lattice vectors for an fcc lattice reads
\begin{align}
    \vb{G} = \frac{2\pi}{a_0} \begin{pmatrix} -l+m+n \\ l-m+n \\ l+m-n \end{pmatrix} ~.
\end{align}
In total, this results in 14 non-vanishing reciprocal lattice vectors that fulfill the condition in Eq.~\eqref{eq:max_G} and have to be considered for umklapp processes.
\section{\label{app:phonon} Phonon-Assisted Densities} 
The full electron-phonon dynamics resulting from the interaction Hamiltonian in Eq.~\eqref{eq:hamiltonian} reads
\begin{align}
    \partial_t f_\mathbf{k}(t)\vert_{scatt}  \label{eq:full_eom_eph} \\
    = -\frac{i}{\hbar} &\sum_{\mathbf{q_1}, \vb{G}, , \pm} g_{\vb{q_1} }^{\vb{G}} \Big(\mp s_{\vb{q_1} ,\vb{k_2}, \vb{k_1}}(t) \mp \Tilde{s}_{-\vb{q_1} ,\vb{k_2}, \vb{k_1}}(t)\Big) ~.  \nonumber 
\end{align}
The quantities $s_{\vb{q_1},\vb{k_2},\vb{k_1}}(t) = \langle b_{\mathbf{q_1}} a^\dagger_\mathbf{k_2} a_\mathbf{k_1} \rangle (t)$ and $\Tilde{s}_{\vb{q_1},\vb{k_2},\vb{k_1}}(t) = \langle b^{\dagger}_{\mathbf{q_1}} a^\dagger_\mathbf{k_2} a_\mathbf{k_1} \rangle (t)$ are phonon-assisted electronic densities describing phonon absorption and emission processes with respect to the corresponding electronic intraband transitions under momentum conservation. Their dynamics are calculated again via a Heisenberg equation of motion involving the single-particle Hamiltonian for electrons and phonons and their interaction and result for phonon absorption in
\begin{align}
    &\Dot{s}_{\vb{q_1},\vb{k_2},\vb{k_1}} = -\frac{i}{\hbar} \Big\{ (\epsilon_{\vb{k_1}} -\epsilon_{\vb{k_2}} + \hbar \omega_{\vb{q_1}}) s_{\vb{q_1},\vb{k_2},\vb{k_1}} \nonumber \\
    &+ \sum_{\vb{q_2},\vb{G'},} g_{-\vb{q_2}}^{\vb{G'}} \langle (1 + b^\dagger_{-\vb{q_2}} b_{\vb{q_1}} + b_{\vb{q_2}} b_{\vb{q_1}}) a^\dagger_\mathbf{k_2} a_{\vb{k_1}-\vb{q_2}-\vb{G'}} \rangle \nonumber \\
    &+ \sum_{\vb{q_2},\vb{G'},}  g_{-\vb{q_2}}^{\vb{G'}} \langle (b^\dagger_{-\vb{q_2}} b_{\vb{q_1}} + b_{\vb{q_2}} b_{\vb{q_1}}) a^\dagger_{\vb{k_2}+\vb{q_2}+\vb{G'}} a_{\vb{k_1}} \rangle \nonumber \\
    &- \sum_{\vb{k'},\vb{G'}} g_{-\vb{q_1}}^{\vb{G'}} \langle a^\dagger_{\vb{k'}-\vb{q_1}+\vb{G'}} a^\dagger_\mathbf{k_2} a_\mathbf{k'} a_\mathbf{k_1} \rangle ~ \label{eq:dt_s}
\end{align}
and for phonon emission in
\begin{align}
    &\Dot{\Tilde{s}}_{\vb{q_1},\vb{k_2},\vb{k_1}} = -\frac{i}{\hbar} \Big\{ (\epsilon_{\vb{k_1}} -\epsilon_{\vb{k_2}} - \hbar \omega_{\vb{q_1}}) \Tilde{s}_{\vb{q_1},\vb{k_2},\vb{k_1}} \nonumber \\
    &+ \sum_{\vb{q_2},\vb{G'},} g_{\vb{q_2} }^{\vb{G'}} \langle (b^\dagger_{\vb{q_1}} b_{\vb{q_2}} + b^\dagger_{\vb{q_1}} b^\dagger_{-\vb{q_2}}) a^\dagger_\mathbf{k_2} a_{\vb{k_1}-\vb{q_2}-\vb{G'}} \rangle \nonumber \\
    &+ \sum_{\vb{q_2},\vb{G'},} g_{\vb{q_2} }^{\vb{G'}} \langle (1 + b^\dagger_{\vb{q_1}} b_{\vb{q_2}} + b^\dagger_{\vb{q_1}} b^\dagger_{-\vb{q_2}}) a^\dagger_{\vb{k_2}+\vb{q_2}+\vb{G'}} a_{\vb{k_1}} \rangle \nonumber \\
    &- \sum_{\vb{k'},\vb{G'}} g_{\vb{q_1} }^{\vb{G'}} \langle a^\dagger_{\vb{k'}+\vb{q_1}+\vb{G'}} a^\dagger_\mathbf{k_2} a_\mathbf{k'} a_\mathbf{k_1} \rangle ~. \label{eq:dt_stilde}
\end{align}
The breakdown of the occurring hierarchy problem is treated in second order Born approximation 
\begin{equation}
    \langle (b^\dagger_{\mathbf{q_1}} b_{\mathbf{q_2}} + b^{(\dagger)}_{\mathbf{q_1}} b^{(\dagger)}_{\mathbf{q_2}}) a^\dagger_\mathbf{k_2} a_\mathbf{k_1} \rangle \approx \langle b^\dagger_{\mathbf{q_1}} b_{\mathbf{q_2}} \rangle \delta_{\vb{q_1},\vb{q_2}} \langle a^\dagger_\mathbf{k_2} a_\mathbf{k_1} \rangle 
\end{equation}
and in a Hartree-Fock like correlation expansion for the four electron operator expectation values \cite{fricke_transport_1996} by neglecting all higher order many-body correlation functions $\langle b^\dagger_{\mathbf{q_1}} b_\mathbf{q_2} a^\dagger_\mathbf{k_2} a_\mathbf{k_1} \rangle^c(t) $ as well as coherent phonon contributions $\langle b^{(\dagger)}_{\mathbf{q_1}} \rangle(t)$. Here, the phonon occupation $n_{\mathbf{q_1}} = \langle b^\dagger_{\mathbf{q_1}} b_{\mathbf{q_1}} \rangle$ is introduced. 

By substituting the momenta from Eq.~\eqref{eq:full_eom_eph} together with a subsequent formal integration of Eq.~(\ref{eq:dt_s}, \ref{eq:dt_stilde}) lead to the integral expressions in Eq.~\eqref{eq:eom_eph}. There, to obtain a local scattering theory for the electronic occupations $f_\mathbf{k}$, terms which are non-diagonal in the electronic momenta are neglected.
In the Markovian limit, these integrals are solved by identifying energy conserving Dirac-$\delta$-distributions in the integrand \cite{haug_quantum_2008}. This results in the Boltzmann scattering equation \eqref{eq:scattering_eq}.
\section{\label{app:drude} Relaxation Time Approximation \& Drude Model}
The Drude model is a classical description for the linear optical response of the quasi-free conduction band electrons of noble metals under the assumption of a phenomenologically introduced Drude collision rate $\gamma_D$ \cite{drude_zur_1900, maier_plasmonics_2007}. In a relaxation time approximation
\begin{align}
    \partial_t f_\mathbf{k}(t) = \frac{e}{\hbar} \vb{E}^{tot}(t) \cdot \nabla_\mathbf{k} f_\mathbf{k}(t) - \gamma_D (f_\mathbf{k}(t) - f_\mathbf{k}^0) ~, \label{eq:rta_boltzmann}
\end{align}
the Drude model is derived from the microscopic dynamics described by the thin film Boltzmann equation \eqref{eq:thin_film_boltzmann}. To derive an expression for the Drude collision time $\gamma_D^{-1}$ from the full electron-phonon scattering equation \eqref{eq:scattering_eq}, the electron distribution $f_\mathbf{k}$ is expanded in linear order with
\begin{align}
    f_\mathbf{k}(t) = f_\mathbf{k}^{eq} + f_\mathbf{k}^1(t) ~, \label{eq:linearization_prescription}
\end{align}
where 
\begin{align}
    f_\mathbf{k}^{eq} = \frac{1}{\exp{\frac{\epsilon_\mathbf{k}-E_F}{k_B T_{eq}}}+1} \label{eq:fermi_dirac} 
\end{align}
is the equilibrium Fermi-Dirac distribution and $f_\mathbf{k}^1(t)$ is the small perturbation from the equilibrium. Due to the conservation of the total electron density, the perturbation term has to fulfill the sum rule \cite{hess_maxwell-bloch_1996-1}
\begin{align}
    \sum_\mathbf{k} f_\mathbf{k}^1(t) = 0 ~. \label{eq:sum_rule}
\end{align}
In accordance with Ref.~\cite{smith_frequency_1982, mahan_solid_2000, czycholl_solid_2023, ziman_electrons_1960}, the linearization results for normal processes in the well-known Bloch-Grüneisen formula for the electrical resistance, where the relaxation rate is given by
\begin{align}
    \gamma_{k}^{N}(T_{eq}) &= \frac{V}{2\pi} \frac{m}{\hbar^3 k_B T_{eq}} \int_0^{q_D} \mathrm{d}q~  \frac{q^3}{k^3} \Phi_{q} |g_{\vb{q}}^{\mathbf{G}=0}|^2 ~, \label{eq:bloch_grüneisen} \\
    \Phi_{q }(T_{eq}) &= \hbar \omega_{q} \frac{1}{ e^{\frac{\hbar \omega_{q}}{k_B T_{eq}}} -1}  \frac{1}{ 1 - e^{-\frac{\hbar \omega_{q}}{k_B T_{eq}}} } ~,
\end{align}
with the Debye wave number $q_{D}$ as maximum transferred phonon momentum in a reduced zone scheme. The remaining $q$-integration is solved numerically.
\\%
The relaxation via umklapp processes requires a more detailed proceeding, since the reciprocal lattice vectors $\vb{G}$ break the isotropy of the $q$-integration. However, by substituting $p=|\vb{q+G}|$ \cite{lawrence_umklapp_1972} and averaging over all directions of the momentum $\vb{k}$ \cite{ziman_electrons_1960}, one obtains
\begin{align}
    \gamma_{\vb{k}}^{U}(T_{eq}) =& \frac{V}{2\pi} \frac{m}{\hbar^3 k_B T_{eq}} \sum_{\vb{G} \neq 0} \int_{G-2k_F}^{q_D} \mathrm{d}q~ \frac{q}{2Gk^3} \Phi_{q} \times \nonumber \\
    &\times \int_{G-q}^{2k_F} \mathrm{d}p~ p^2 |g_\mathbf{q}^\mathbf{G=p-q}|^2~.
\end{align}
Here, the boundaries of the integrals are important to include only scattering processes on the Fermi surface obeying the condition in Eq.~\eqref{eq:max_G}. Thus, the set of possible reciprocal lattice vectors is limited to those with the shortest length \cite{smith_frequency_1982}, App.\ref{app:umklapp}. Inserting the special electron-phonon matrix element applied in this work (cf.~Eq.~\eqref{eq:def_eph_coupling}) and carrying out the $p$-integration yields
\begin{align}
    &\gamma_{k}^{U}(T_{eq}) = \frac{V}{2\pi} \frac{m}{\hbar^3 k_B T_{eq}} \sum_{\vb{G} \neq 0} \int_{G-2k_F}^{q_D} \mathrm{d}q~ \frac{q}{2Gk^3} \Phi_{q} \times \\
    & \times \frac{e^2 \hbar \omega_q}{2 V \varepsilon_0} \Big( 2k_F -G+q - \kappa \arctan{\Big[ \frac{\kappa(2k_F-G+q)}{\kappa^2 + 2k_F(G-q)} \Big] } \Big) ~. \nonumber
\end{align}
In agreement with our numerical results for the orientational relaxation time from Sec.~\ref{sec:results}, the Drude collision time is estimated at the Fermi edge with $\gamma_D^{-1} = (\gamma^{N}_{k_F} + \gamma^{U}_{k_F})^{-1} = 19.9$~fs, where the single contributions are $(\gamma^{N}_{k_F})^{-1} = 83.0$~fs and $(\gamma^{U}_{k_F})^{-1} = 26.1$~fs.
\\%  
To derive the Drude model, the linearization from Eq.~\eqref{eq:linearization_prescription} is introduced to Eq.~\eqref{eq:rta_boltzmann} and solved in the frequency domain
\begin{align}
    \hat{f}_\mathbf{k}^1(\omega) = - \frac{\frac{e}{\hbar} \hat{\vb{E}}^{tot}(\omega) \cdot \nabla_\mathbf{k} f_\mathbf{k}^{eq} }{ i\omega - \gamma_D } ~.
\end{align}
The macroscopic Drude model results from the current density calculated with Eq.~(\ref{eq:coarse_grained},\ref{eq:susceptibility}). For a parabolic dispersion integration by parts with vanishing boundary terms solves $\sum_\mathbf{k} \vb{v_k} \nabla_\mathbf{k} f_\mathbf{k} = - \frac{\hbar}{m} \mathds{1} \sum_\mathbf{k} f_\mathbf{k}$ and results in the Drude susceptibility
\begin{align}
    \chi_D(\omega) = \Tilde{\varepsilon}_b - \frac{\omega_p^2}{\omega^2 + i \omega \gamma_D} \label{eq:chi_drude}
\end{align}
with the plasma frequency $\omega_p^2= \frac{e^2n_0}{\varepsilon_0 m}$. Here, we add an effective dielectric background via $\Tilde{\varepsilon}_b = \varepsilon_b-\varepsilon_{out}$ phenomenologically as in Eq.~\eqref{eq:source_current}. 
\section{\label{app:induced_field} Induced Field in the Thin Film}
As restriction of the formal solution in Eq.~\eqref{eq:formal_solution} to the special thin film geometry shown in Fig.~\ref{fig:thin_film}, the incidence of the irradiated field $\vb{E}_0$ is assumed in $z$-direction perpendicular to the thin film with polarization in $y$-direction. 
Assuming a purely in-plane current density leads to a reduced one-dimensional problem for the solution of the wave equation. The current density $j_s(t)$ localized in the thin film between $z=0$ and $z=d$
\begin{equation}
    \vb{j}_s(\vb{r},t) = \Theta(z) \big(1-\Theta(z-d) \big) j_s(t) \vb{e}_y
\end{equation}
is described as spatial homogeneous in the thin film via the Heaviside-$\Theta$ function. Thus, the inhomogeneous solution of the wave equation is solved in cylindrical coordinates
\begin{align}
    &\vb{E}_{ind}(z,t) = \partial_t \int \mathrm{d}^3r'~ \frac{j_s \Big(z', t \pm \frac{\vert \vb{r}-\vb{r}'\vert}{c} \Big) \vb{e_y}}{\vert \vb{r}-\vb{r}' \vert} \nonumber\\
    &= 2\pi \partial_t \int_{-\infty}^\infty \mathrm{d}z'~ \Theta(z') \big(1-\Theta(z'-d)) \times \nonumber \\
    & ~~~~~\times \int_0^\infty \mathrm{d}\rho' \rho'~\frac{j_s \Big(t \pm \frac{\sqrt{(z-z')^2+\rho'^2}}{c} \Big)\vb{e_y}}{\sqrt{(z-z')^2+\rho'^2}} \nonumber \\
    &= 2\pi \partial_t \int_{0}^{d} \mathrm{d}z'~ \int_{\frac{\vert z-z'\vert}{c}}^\infty \mathrm{d}\xi~ j_s(t \pm \xi) \vb{e_y} ~.
\end{align}
Here, the substitution $\xi = \frac{\sqrt{(z-z')^2+\rho'^2}}{c}$ is applied. Since the current density depends now just on the retarded time $t\pm \xi$, we can rewrite the time derivative via a derivative by $\xi$ and solve the $\xi$-integral
\begin{align}
    &= \mp 2\pi \int_{0}^{d} \mathrm{d}z'~ \int_{\frac{\vert z-z'\vert}{c}}^\infty \mathrm{d}\xi~ \partial_\xi j_s(t \pm \xi) \vb{e_y}\nonumber \\
    &= \pm 2\pi \int_{0}^{d} \mathrm{d}z'~ j_s \Big(t \pm \frac{\vert z-z'\vert}{c} \Big) \vb{e_y} ~.
\end{align}
In the approximation of a small film thickness compared to the wavelength $d\ll \lambda_L$, retardation effects in the film are negligible. Finally, this results in the inhomogeneous solution of the wave equation in Eq.~\eqref{eq:induced_field}.
\\%
\textit{Radiative damping:~} The dynamics of the macroscopic current density obtained from Eq.~\eqref{eq:thin_film_boltzmann} is affected by a self-interaction via the re-emitted field and the background polarization
\begin{align}
    \partial_t \vb{j} \propto& -\frac{e^2}{\hbar} \frac{2}{V} \sum_\mathbf{k} \vb{v_k}\cdot \nabla_\mathbf{k} f_\mathbf{k} \vb{E}_{ind}(\vb{j},\vb{P}_b) ~.
\end{align}
Applying the induced field from Eq.~\eqref{eq:induced_field} and solving in frequency domain yields a radiative damping rate $\gamma_{rad}$
\begin{align}   
    -i\omega \hat{\vb{j}} \propto - \frac{\omega_p^2 d}{2c_0 n_{out}} \frac{1+i\omega (\varepsilon_{out} -\varepsilon_b) \frac{d}{c_0 n_{out}}}{1+i\omega (\varepsilon_{out}-\varepsilon_b) \frac{d}{2c_0 n_{out}}} \hat{\vb{j}} = - \gamma_{rad} \hat{\vb{j}}~, \label{eq:derivation_gamma_rad}
\end{align}
which depends only on the total electron density $n_0 = \frac{2}{V} \sum_\mathbf{k} f_\mathbf{k}$. Therefore, since the density has to be conserved for all times in a spatial homogeneous system, the self-interaction does not contribute non-linearities to the macroscopic current density.
\bibliography{references}
\end{document}